\documentclass[10pt]{iopart}

\usepackage{graphicx}
\usepackage{dcolumn}
\usepackage{bm}
\usepackage{color}

\begin{document}

\title[NBI current drive]{Effect of the tangential NBI current drive on the stability of pressure and energetic particle driven MHD modes in LHD plasma}


\author{J. Varela}
\ead{jacobo.varela@nifs.ac.jp}
\address{National Institute for Fusion Science, National Institute of Natural Science, Toki, 509-5292, Japan}
\author{W. A. Cooper}
\address{Ecole Polytechnique de Lausanne (EPFL), Centre de Recherches en Physique des Plasma (CRPP), CH-1015 Lausanne, Switzerland}
\author{K. Nagaoka}
\address{National Institute for Fusion Science, National Institute of Natural Science, Toki, 509-5292, Japan}
\author{K. Y. Watanabe}
\address{National Institute for Fusion Science, National Institute of Natural Science, Toki, 509-5292, Japan}
\author{D. A. Spong}
\address{Oak Ridge National Laboratory, Oak Ridge, Tennessee 37831-8071, USA}
\author{L. Garcia}
\address{Universidad Carlos III de Madrid, 28911 Leganes, Madrid, Spain}
\author{A. Cappa}
\address{Laboratorio Nacional de Fusion CIEMAT, Madrid, Spain}
\author{A. Azegami}
\address{National Institute for Fusion Science, National Institute of Natural Science, Toki, 509-5292, Japan}

\date{\today}

\begin{abstract}
The aim of the present study is to analyze the stability of the pressure gradient driven modes (PM) and Alfv\'en eigenmodes (AE) in the Large Helical Device (LHD) plasma if the rotational transform profile is modified by the current drive of the tangential neutral beam injectors (NBI). This study forms a basic search for optimized operation scenarios with reduced mode activity. The analysis is performed using the code FAR3d which solves the reduced MHD equations describing the linear evolution of the poloidal flux and the toroidal component of the vorticity in a full 3D system, coupled with equations for density and parallel velocity moments of the energetic particle (EP) species, including the effect of the acoustic modes. The Landau damping and resonant destabilization effects are added via the closure relation. On-axis and off-axis NBI current drive modifies the rotational transform which becomes strongly distorted as the intensity of the neutral beam current drive (NBCD) increases, leading to wider continuum gaps and modifying the magnetic shear. The simulations with on-axis NBI injection show that a counter (ctr-) NBCD in inward shifted and default configurations leads to a lower growth rate of the PM, although strong $n=1$ and $2$ AEs can be destabilized. For the outward shifted configurations, a co-NBCD improves the AEs stability but the PM are further destabilized if the co-NBCD intensity is $30$ kA/T. If the NBI injection is off-axis, the plasma stability is not significantly improved due to the further destabilization of the AE and energetic particle modes (EPM) in the middle and outer plasma region.
\end{abstract}

%
%
%
%
%


\vspace{2pc}

\maketitle

\ioptwocol

\section{Introduction \label{sec:introduction}}

The magnetic field topology of nuclear fusion devices is modified if non inductive currents are generated in the plasma. Non inductive currents can be self generated as in the case of the bootstrap current, which is driven by the collisions between trapped and passing particles \cite{1,2,3}, or generated by the external injection of lower Hybrid waves (LHW) \cite{4,5}, electron cyclotron waves (ECW) \cite{6,7} and neutral beams (NBI) \cite{8,9}.

The non inductive current drive is a promising mechanism to achieve steady state operation in advanced tokamaks where large bootstrap currents replace the magnetic field component generated by the transformer coils \cite{10,11,12}. In addition, the non inductive current drive is used to modify the magnetic field configuration of the fusion devices, for example by the electron cyclotron current drive (ECCD) \cite{13,14} and the neutral beam current drive (NBCD) \cite{15,16,17,18}, leading to an improved stability of the pressure and current gradient driven modes (PM) \cite{19,20,21,22,23,24} as well as the Alfv\'en Eigenmodes (AE)\cite{25,26}. ECCD is also used in stellarators \cite{27,28,29,30} to improve the stability properties of the plasma with respect to the PM and AE \cite{31,32,33,34,35,36,37,38,39}. In particular, the effect of the ECCD and the NBCD was analyzed on LHD plasma \cite{40,41} attaining the stabilization of the energetic-ion-driven resistive interchange mode (EIC) \cite{42,43,44}, Toroidal and global Alfv\'en eigenmodes (TAE / GAE) \cite{45} as well as PM \cite{46,47}.

The PM limits the performance of LHD plasma causing a transport degradation and lower confinement \cite{48,49,50,51,52}. Also, the energetic particle (EP) driven instabilities enhance the transport of fusion produced alpha particles (in ignited devices), energetic hydrogen neutral beams and ion cyclotron resonance heated particles (ICRF) \cite{53,54,55}, producing a decrease of the heating efficiency in helical devices such as LHD and W7-AS stellarators or tokamaks such as JET and DIII-D \cite{55,57,58,59,60,61,62,63,64}. The EP losses are triggered if there is a resonance between the unstable mode frequency and the EP drift, bounce or transit frequencies.

\begin{figure}[h!]
\centering
\includegraphics[width=0.45\textwidth]{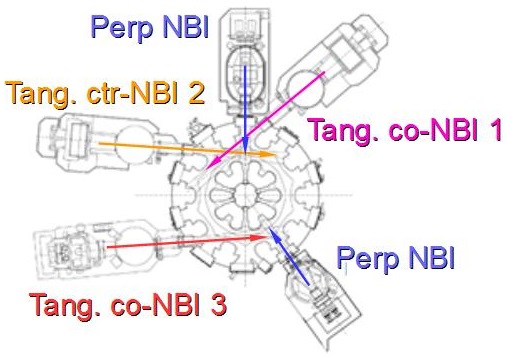}
\caption{Schematic view of the NBI lines in LHD.}\label{FIG:1}
\end{figure}

LHD is a helical device heated by three tangential NBI lines parallel to the magnetic axis and deposited in the plasma core with a beam energy of 180 keV. In addition, two NBIs perpendicular to the magnetic axis injected in the plasma periphery with a beam energy of 32 keV \cite{65,66,67}. Figure~\ref{FIG:1} shows a schematic view of the NBI injection lines in LHD. The tangential NBIs are oriented in the clockwise and counter-clockwise directions to balance the current generated by the beams, although depending on the NBIs operational regime a net plasma current ($I_{p}$) can be generated during the discharge, modifying the magnetic field configuration, particularly the rotational transform profile. Consequently, the stability of the PM and the AE changes.

\begin{figure*}[h!]
\centering
\includegraphics[width=0.9\textwidth]{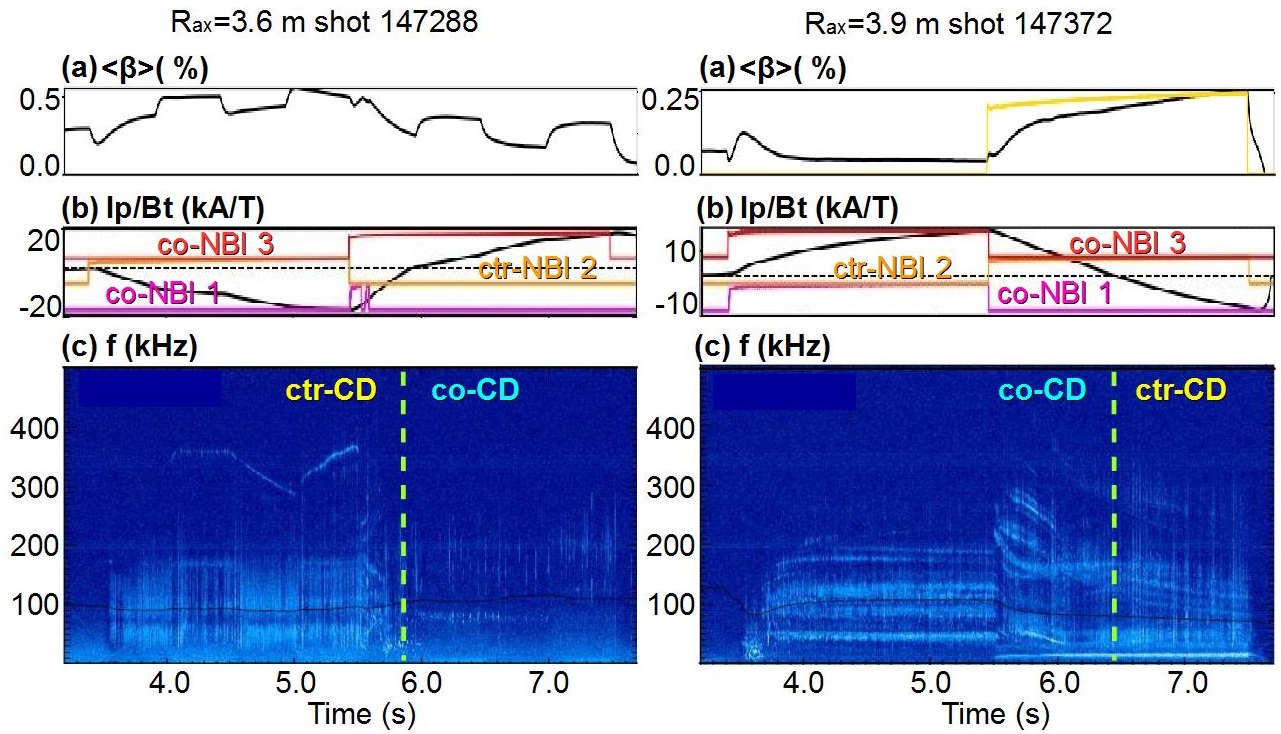}
\caption{LHD discharge with $R_{ax} = 3.6$ m (shot 147288) and with $R_{ax} = 3.9$ m (shot 147372). Panel (a) shows the averaged $\beta$, (b) toroidal current and NBI injection pattern and (c) raw signal of the magnetic probe.}\label{FIG:2}
\end{figure*}

The aim of the present study is to identify optimized operation scenario with reduced mode activity through analyzing the stability of PM and AE in LHD configurations with different locations of the vacuum magnetic axis locations ($R_{ax}$), with respect to the net plasma current generated by the tangential NBIs. In addition, the effect of the NBI deposition region on the magnetic field configuration is included in the analysis identifying optimization trends for an off-axis beam injection. Figure~\ref{FIG:2}c shows a spectrogram of the time evolution of the magnetic probe signal during the discharges $147288$ ($R_{ax} = 3.6$ m and $B= 1.375$ T) and $147372$ ($R_{ax} = 3.9$ m and $B= 1.375$ T). The simulations will reproduce the trends observed in the co- and ctr-NBCD phases of the discharges, for example, the increase of the AE families frequency ranges and the weaker magnetic probe signal during the co-NBCD phase with respect to the ctr-NBCD phase as well as the  further destabilization of low (high) frequency AEs during the co-(ctr-) NBCD phase.

The simulations are performed using the gyro-fluid code FAR3d \cite{68}, which is an extended version of the original FAR3d code that solves the reduced linear resistive MHD equations \cite{69,70,71}, adding the moment equations of the energetic ion density and parallel velocity \cite{73,74}. The model reproduces the linear wave-particle resonance effects required for Landau damping/growth and the parallel momentum response of the thermal plasma required for coupling to the geodesic acoustic waves \cite{75}. The simulations are based on an equilibria calculated by VMEC code \cite{72}.

The main motivation of performing the analysis with the gyro-fluid code FAR3d is the computational efficiency; this is due to its reduction of selected kinetic effects to a set of 3D fluid-like equations rather than more complex approaches, for example initial value gyrokinetic codes as EUTERPE \cite{76}, GEM \cite{77}, GYRO \cite{78}, GTC \cite{79}, ORB5 \cite{80} and GENE \cite{107} or kinetic-MHD hybrid codes as MEGA \cite{81}. FAR3d can be used for rapid parameter/profile scans in order to perform optimization/design studies where rapidly evaluated physics target functions are required. Also, the code can be used to identify AE stability trends of EP instabilities since critical fast ion characteristics, such as the density profile often cannot directly be measured. It should be noted that the Landau closure model used in FAR3d code is the only known non perturbative energetic particle stability model where it is feasible to do an eigenmode analysis (GYRO and GENE have the capability of eigensolver analysis although the matrix are too large for a widespread use). Finally, in comparison to particle-based methods, this approach has the advantages of zero noise levels, exact implementation of boundary conditions and an improved ability to include extended mode coupling effects. On the other hand, the simplification of the kinetic effects can lead to a deviation of FAR3d results compared to more complete approaches, although a methodology has been developed for calibrating the Landau-closure against more complete kinetic models through optimization of the closure coefficients \cite{75}. A detail comparison between codes was recently performed by other authors \cite{82}.

This paper is organized as follows. The model equations, numerical scheme and equilibrium properties are described in section \ref{sec:model}. The stability analysis of pressure gradient driven modes and Alfv\'en Eigenmodes triggered by a NBI deposited on-axis, with respect to the magnetic axis location and the NBCD intensity, is performed in section \ref{sec:NBI_onaxis}. Next, the effect of off-axis NBI deposition is studied in section \ref{sec:NBI_offaxis}. The comparison of the theoretical results and experimental data is performed in section \ref{sec:exp}. Finally, the conclusions of this paper are presented in section \ref{sec:conclusions}.

\section{Equations and numerical scheme \label{sec:model}}

Following the method employed in Ref.\cite{83}, a reduced set of equations for high-aspect ratio configurations and moderate $\beta$-values (of the order of the inverse aspect ratio) is derived retaining the toroidal angle variation, based upon an exact three-dimensional equilibrium that assumes closed nested flux surfaces. The effect of the energetic particle population on the plasma stability is included through moments of the fast ion kinetic equation truncated with a closure relation \cite{84}, describing the evolution of the perturbation of the energetic particle density ($\tilde n_{f}$) and velocity moments parallel to the magnetic field lines ($\tilde v_{||f}$). The coefficients of the closure relation are selected to match analytic TAE growth rates based upon a two-pole approximation of the plasma dispersion function (Maxwellian EP distribution). All functions have equilibrium and perturbation components represented as: $ A = A_{eq} + \tilde{A} $.

The model formulation assumes high aspect ratio, medium $\beta$ (of the order of the inverse aspect ratio $\varepsilon=a/R_0$), small variation of the fields and small resistivity. The plasma velocity and perturbation of the magnetic field are defined as
\begin{equation}
 \mathbf{v} = \sqrt{g} R_0 \nabla \zeta \times \nabla \Phi, \quad\quad\quad  \mathbf{\tilde{B}} = R_0 \nabla \zeta \times \nabla \tilde\psi,
\end{equation}
where $\zeta$ is the toroidal angle, $\Phi$ is a stream function proportional to the electrostatic potential, and $\tilde \psi$ is the perturbation of the poloidal flux.

The model equations include the time evolution of the perturbations of the poloidal flux, the toroidal component of the vorticity ($\tilde U$), the pressure ($\tilde p$), the parallel velocity of the thermal particles ($v_{||th}$), the EP density ($\tilde n_f$) and the EP parallel velocity ($\tilde v_{\| f}$). The equations, in dimensionless form, are
\begin{equation}
\frac{\partial \tilde \psi}{\partial t} =  \sqrt{g} B \nabla_\| \Phi  + \frac{\eta}{S} \tilde J^\zeta
\end{equation}
\begin{eqnarray} 
\frac{{\partial \tilde U}}{{\partial t}} =  - v_{\zeta,eq} \frac{\partial \tilde U}{\partial \zeta} \nonumber\\
+ \sqrt{g} B  \nabla_{||} \tilde J^{\zeta} - \frac{1}{\rho} \left( \frac{\partial J^{\zeta}_{eq}}{\partial \rho} \frac{\partial \tilde \psi}{\partial \theta} - \frac{\partial J^{\zeta}_{eq}}{\partial \theta} \frac{\partial \tilde \psi}{\partial \rho}    \right) \nonumber\\
- {\frac{\beta_0}{2\varepsilon^2} \sqrt{g} \left( \nabla \sqrt{g} \times \nabla \tilde p \right)^\zeta } -  {\frac{\beta_f}{2\varepsilon^2} \sqrt{g} \left( \nabla \sqrt{g} \times \nabla \tilde n_f \right)^\zeta }
\end{eqnarray} 
\begin{eqnarray}
\label{pressure}
\frac{\partial \tilde p}{\partial t} = - v_{\zeta,eq} \frac{\partial \tilde p}{\partial \zeta} + \frac{dp_{eq}}{d\rho}\frac{1}{\rho}\frac{\partial \tilde \Phi}{\partial \theta} \nonumber\\
 +  \Gamma p_{eq}  \left[{\left( \nabla \sqrt{g} \times \nabla \tilde \Phi \right)^\zeta - \nabla_\|  \tilde v_{\| th} }\right] 
\end{eqnarray} 
\begin{eqnarray}
\label{velthermal}
\frac{{\partial \tilde v_{\| th}}}{{\partial t}} = - v_{\zeta,eq} \frac{\partial \tilde v_{||th}}{\partial \zeta} -  \frac{\beta_0}{2n_{0,th}} \nabla_\| p 
\end{eqnarray}
\begin{eqnarray}
\label{nfast}
\frac{{\partial \tilde n_f}}{{\partial t}} = - v_{\zeta,eq} \frac{\partial \tilde n_{f}}{\partial \zeta} - \frac{v_{th,f}^2}{\varepsilon^2 \omega_{cy}}\ \Omega_d (\tilde n_f) - n_{f0} \nabla_\| \tilde v_{\| f}   \nonumber\\
-  n_{f0} \, \Omega_d (\tilde \Phi) + n_{f0} \, \Omega_* (\tilde  \Phi) 
\end{eqnarray}
\begin{eqnarray}
\label{vfast}
\frac{{\partial \tilde v_{\| f}}}{{\partial t}} = - v_{\zeta,eq} \frac{\partial \tilde v_{||f}}{\partial \zeta}  -  \frac{v_{th,f}^2}{\varepsilon^2 \omega_{cy}} \, \Omega_d (\tilde v_{\| f}) \nonumber\\
- \left( \frac{\pi}{2} \right)^{1/2} v_{th,f} \left| \nabla_\| \tilde v_{\| f}  \right| - \frac{v_{th,f}^2}{n_{f0}} \nabla_\| n_f + v_{th,f}^2 \, \Omega_* (\tilde \psi) 
\end{eqnarray}
Equation (2) is derived from Ohm$'$s law coupled with Faraday$'$s law, equation (3) is obtained from the toroidal component of the momentum balance equation after applying the operator $\nabla \wedge \sqrt{g}$, equation (4) is obtained from the thermal plasma continuity equation with compressibility effects and equation (5) is obtained from the parallel component of the momentum balance \cite{69,70,71,83}. Equations (6) and (7) are obtained calculating the first two moments of the kinetic equation \cite{74,75}. Here, $\tilde U =  \sqrt g \left[{ \nabla  \times \left( {\rho _m \sqrt g {\bf{v}}} \right) }\right]^\zeta$ is the toroidal component of the vorticity, $\rho_m$ the ion and electron mass density, $\rho = \sqrt{\phi_{N}}$ the effective radius with $\phi_{N}$ the normalized toroidal flux and $\theta$ the poloidal angle. The perturbation of the toroidal current density $\tilde J^{\zeta}$ is defined as:
\begin{eqnarray}
\tilde J^{\zeta} =  \frac{1}{\rho}\frac{\partial}{\partial \rho} \left(-\frac{g_{\rho\theta}}{\sqrt{g}}\frac{\partial \tilde \psi}{\partial \theta} + \rho \frac{g_{\theta\theta}}{\sqrt{g}}\frac{\partial \tilde \psi}{\partial \rho} \right) \nonumber\\
- \frac{1}{\rho} \frac{\partial}{\partial \theta} \left( -\frac{g_{\rho\rho}}{\sqrt{g}}\frac{1}{\rho}\frac{\partial \tilde \psi}{\partial \theta} + \rho \frac{g_{\rho \theta}}{\sqrt{g}}\frac{\partial \tilde \psi}{\partial \rho} \right)
\end{eqnarray}
and the toroidal component of the equilibrium current density is:
\begin{equation}
J^{\zeta}_{eq} = \frac{1}{\varepsilon^2} \frac{1}{\rho} \frac{dI}{d\rho} - \frac{\partial \beta_{*}}{\partial \theta}
\end{equation}
$v_{\zeta,eq}$ is the equilibrium toroidal rotation and $I$ is the toroidal current. $\beta_{0}$ is the equilibrium $\beta$ at the magnetic axis, $\beta_{f0}$ is the EP $\beta$ at the magnetic axis and $n_{f0}$ is the EP radial density profile normalized to its value at the magnetic axis. $\Phi$ is normalized to $a^2B_{0}/\tau_{A0}$ and $\tilde\psi$ to $a^2B_{0}$ with $\tau_{A0}$ the Alfv\' en time $\tau_{A0} = R_0 (\mu_0 \rho_m)^{1/2} / B_0$. The radius $\rho$ is normalized to a minor radius $a$; the resistivity to $\eta_0$ (its value at the magnetic axis); the time to the Alfv\' en time; the magnetic field to $B_0$ (the averaged value at the magnetic axis); and the pressure to its equilibrium value at the magnetic axis. The Lundquist number $S$ is the ratio of the resistive time $\tau_{R} = a^2 \mu_{0} / \eta_{0}$ to the Alfv\' en time. $\rlap{-} \iota$ is the rotational transform, $v_{th,f} = \sqrt{T_{f}/m_{f}}$ is the radial profile of the energetic particle thermal velocity normalized to the Alfv\' en velocity at the magnetic axis $v_{A0}$ and $\omega_{cy}$ the energetic particle cyclotron frequency normalized to $\tau_{A0}$. $q_{f}$ is the charge, $T_{f}$ is the radial profile of the effective EP temperature and $m_{f}$ is the mass of the EP. The $\Omega_{d}$ and $\Omega_{*}$ operators are defined as:
\begin{eqnarray}
\label{eq:omedrift}
\Omega_d = \frac{1}{2 B^4 \sqrt{g}}  \Bigg\{ \left( \frac{I}{\rho} \frac{\partial B^2}{\partial \zeta} - J \frac{1}{\rho} \frac{\partial B^2}{\partial \theta} \right) \frac{\partial}{\partial \rho} \nonumber\\
-  \left( \rho \beta_* \frac{\partial B^2}{\partial \zeta} - J \frac{\partial B^2}{\partial \rho} \right) \frac{1}{\rho} \frac{\partial}{\partial \theta} \nonumber\\ 
+ \left( \rho \beta_* \frac{1}{\rho} \frac{\partial B^2}{\partial \theta} -  \frac{I}{\rho} \frac{\partial B^2}{\partial \rho} \right) \frac{\partial}{\partial \zeta} \Bigg\}
\end{eqnarray}

\begin{eqnarray}
\label{eq:omestar}
\Omega_* = \frac{1}{B^2 \sqrt{g}} \frac{1}{n_{f0}} \frac{d n_{f0}}{d \rho} \left( \frac{I}{\rho} \frac{\partial}{\partial \zeta} - J \frac{1}{\rho} \frac{\partial}{\partial \theta} \right) 
\end{eqnarray}
Here the $\Omega_{d}$ operator is constructed to model the average drift velocity of a passing particle and $\Omega_{*}$ models its diamagnetic drift frequency. These operators, using an approximated Maxwell distribution function, can treat the instabilities driven by passing EP. $J$ is the equilibrium poloidal current. We also define the parallel gradient and curvature operators as
\begin{equation}
\label{eq:gradpar}
\nabla_\| f = \frac{1}{B \sqrt{g}} \left( \frac{\partial \tilde f}{\partial \zeta} - \rlap{-} \iota \frac{\partial \tilde f}{\partial \theta} - \frac{\partial f_{eq}}{\partial \rho}  \frac{1}{\rho} \frac{\partial \tilde \psi}{\partial \theta} + \frac{1}{\rho} \frac{\partial f_{eq}}{\partial \theta} \frac{\partial \tilde \psi}{\partial \rho} \right)
\end{equation}
\begin{equation}
\label{eq:curv}
\sqrt{g} \left( \nabla \sqrt{g} \times \nabla \tilde f \right)^\zeta = \frac{\partial \sqrt{g} }{\partial \rho}  \frac{1}{\rho} \frac{\partial \tilde f}{\partial \theta} - \frac{1}{\rho} \frac{\partial \sqrt{g} }{\partial \theta} \frac{\partial \tilde f}{\partial \rho}
\end{equation}
with the Jacobian of the transformation,
\begin{equation}
\label{eq:Jac}
\frac{1}{\sqrt{g}} = \frac{B^2}{(J- \rlap{-} \iota I)}
\end{equation}
Equations~\ref{pressure} and~\ref{velthermal} introduce the parallel momentum response of the thermal plasma. These are required for coupling to the geodesic acoustic waves, accounting for the geodesic compressibility in the frequency range of the geodesic acoustic mode (GAM) \cite{85,86}.

Equilibrium flux coordinates $(\rho, \theta, \zeta)$ are used. Here, $\rho$ is normalized to the unity at the edge. The flux coordinates used in the code are those described by Boozer \cite{87}.

The FAR3d gyro-fluid code uses finite differences in the radial direction and Fourier expansions in the two angular variables. Two numerical schemes to resolve the linear equations can be used in the code: a semi-implicit initial value or an eigenvalue solver. The initial value solver calculates the mode with the largest growth rate (dominant mode) and the eigen-solver the stable and unstable modes (sub-dominant modes). The analysis of the sub-dominant modes is required to calculate the growth rate of the multiple AE families that can be unstable or marginally unstable during the discharge. In addition, the study of the sub-dominant modes is motivated by the fact that the equilibrium profiles are not known precisely from the experiment. This can result in a more close correspondence of sub-dominant modes with the experimentally observed modes than the fastest growing mode. In this way, the eigenmode can provide an uncertainty characterization both in the modeling and the measurements.

The present model was already used to study the stability of PM during sawtooth-like events and internal collapses \cite{88,89,90,91}, TAE \cite{68} and EIC \cite{92} activity in LHD, indicating reasonable agreement with the observations.

\subsection{Equilibrium properties}

A set of fixed boundary results from the VMEC equilibrium code \cite{72} is calculated including the distortion of the $\rlap{-} \iota$ profile caused by the NBI current drive for an inward shifted ($R_{ax} = 3.5$ m), default ($R_{ax} = 3.6$ m) and outward shifted ($R_{ax} = 3.75$ and $3.9$ m) LHD configurations. Three NBI deposition regions are assumed in the analysis, one on-axis and two off-axis cases, where the beam is injected at the magnetic axis, the middle plasma region or the plasma periphery. In the case with on-axis NBI deposition four NBCD intensities are analyzed from $I_{p} = 0$ kA/T (balanced current case) to $30$ kA/T (co-NBCD cases) for a $\Delta I_{p} = 10$ kA/T. With respect to the cases with off-axis NBI deposition, the NBCD intensities included in this study range between $-10$ kA/T (ctr-NBCD case) to $30$ kA/T for a $\Delta I_{p}=10$ kA/T). The ctr-NBCD cases with an $\rlap{-} \iota$ profile that decreases below $0.2$ are not included in the study, because the VMEC equilibria do not converge or the transformation of the equilibria to the Boozer coordinates leads to a bad conversion of the magnetic surfaces in the inner plasma region, thus the PM/AE stability cannot be analyzed properly in this regime.

Table~\ref{Table:1} shows the main parameters of the Hydrogen thermal plasma and table~\ref{Table:2} the details of the EP injected by the tangential Hydrogen NBI ($<\beta_{f}> = 0.34 \%$ corresponds to a $\beta_{f0} = 1 \%$). The cyclotron frequency is $\omega_{cy} = 9.58 \cdot 10^{7}$ s$^{-1}$.

\begin{table}[t]
\centering
\begin{tabular}{c c c c}
\hline
$T_{th,0}$ & $n_{th,0}$ & $<\beta_{tot}>$ & $V_{A0}$ \\
(keV) & ($10^{20}$ m$^{-3}$) & ($\%$) & ($10^{6}$ m/s) \\ \hline
4 & 0.3 & 1.8 & 3.98 \\
\end{tabular}
\caption{Thermal plasma properties in the reference case (values at the magnetic axis). The first column is the thermal temperature, the second column is the thermal density, the third column is the thermal $\beta$ and the fourth column is the Alfv\' en velocity.} \label{Table:1}
\end{table}

\begin{table}[t]
\centering
\begin{tabular}{c c c c}
\hline
$T_{NBI}$ & $T_{f0}$ & $n_{f,0}$ & $<\beta_{f}>$ \\
(keV) &  (keV) & ($10^{20}$ m$^{-3}$) & ($\%$)\\ \hline
180 & 100 & 0.0025 & 0.34  \\ 
\end{tabular}
\caption{Properties of the EP driven by the tangential NBI in the reference case (maximum value). First column is the NBI injection energy, the second column is the EP energy, the third column is the EP density and the forth column the EP $\beta$.} \label{Table:2}
\end{table}

The magnetic field at the magnetic axis is $1$ T and the averaged inverse aspect ratio extends from $\varepsilon=0.175$ in the default configuration to $0.141$ in the outward shifted equilibria with $R_{ax} = 3.9$ m. 

\begin{figure}[h!]
\centering
\includegraphics[width=0.45\textwidth]{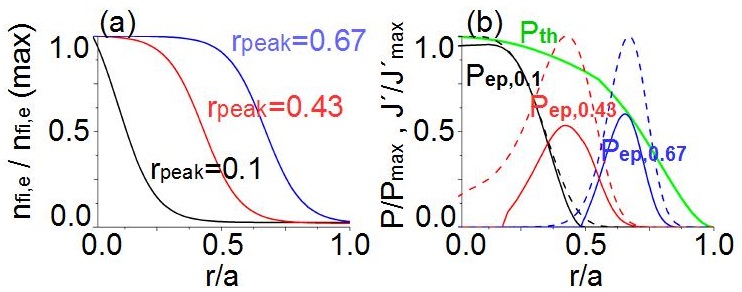}
\caption{EP density in the on-axis (black line) and off axis cases with $r_{peak}=0.43$ (red line) and $r_{peak}=0.67$ (blue line) (a). Thermal pressure (green line) and EP pressure in the on-axis (black line) and off axis cases with $r_{peak}=0.43$ (red line) and $r_{peak}=0.67$ (blue line) (b). The dashed lines show the current density induced by the beam in the on-axis and off axis cases (b). The profiles are normalized to the maximum value.}\label{FIG:3}
\end{figure}

The energy of the injected particles by the tangential NBI is $T_{NBI}(0) = 180$ keV, but we take the nominal EP temperature $T_{f}(0) = 100$ keV ($v_{th,f} = 3.09 \cdot 10^{6}$ m/s), the approximated EP temperature for a Maxwellian EP distribution fitting the slowing down distribution function of a NB with $180$ keV. For simplicity no radial dependency of the EP energy is considered in the study. The EP density profile is given by the analytic expression:

\begin{equation}
\label{EP_dens}
$$n_{f}(r) = \frac{(0.5 (1+ \tanh(\delta_{r} \cdot (r_{peak}-r)))+0.02)}{(0.5 (1+\tanh(\delta_{r} \cdot r_{peak}))+0.02)}$$
\end{equation}
with the location of the EP density gradient defined by the variable $r_{peak}$ and the flatness by $\delta_{r}$. The parameter $r_{peak}$ is set to $0.1$ in the cases with on-axis NBI deposition and takes the values $0.43$ and $0.67$ in the two cases with off-axis NBI deposition. The parameter $\delta_{r}$ is linked to the available free energy to destabilize the AEs, thus a larger $\delta_{r}$ indicates a stronger NBI drive. $\delta_{r}$ is fixed to $7$ in the simulations.  The NBCD is introduced in the model as a Gaussian centered at the deposition region of the NBI, expressed as:
\begin{equation}
\label{eq:Gauss}
f(r) = I_{p,max} e^{-\frac{1}{2}\left(\frac{r - r_{peak}}{\delta_{I}} \right)^2}
\end{equation}
with $I_{p,max}$ the local maxima of the NBCD intensity (between $-10$ to $30$ kA) and $\delta_{I}$ the width of the Gaussian ($0.19$ for the on-axis case, $0.11$ if the NBI deposition is at $r_{peak} = 0.43$ and $0.07$ if the NBI is deposited at $r_{peak} = 0.67$).
Figure~\ref{FIG:3} (a) indicates the normalized EP density for the on-axis and off-axis NBI deposition cases. Figure~\ref{FIG:3} (b) shows the thermal / EP pressure and plasma current for cases with on axis and off-axis NBI injection (the pressure is  normalized to maximum value of the thermal pressure). The EP density profile in the simulations with off-axis NBI injection is flat rather than hollow in the inner plasma. This simplification is used to avoid the triggering of AE instabilities in the inner part of the plasma, focusing the analysis on the AEs destabilized by a single EP density gradient. The contribution of the EP to the total pressure is calculated by the ANIMEC code using a slowing-down distribution for the EP fitted to an isotropic case \cite{93,94}. It should be noted that there is an inconsistency between the simplified profiles of the EP density and temperature used in the study of the AE stability with respect to the EP pressure component included in the calculation of the VMEC equilibria. The EP pressure component used in VMEC does not include the profile simplifications in order to improve the simulation accuracy calculating the stability of the PM. Also, the component of the EP pressure in the total pressure is larger compared to a typical LHD discharge, although this selection facilitates an easy identification of the PM/AE stability trends between simulations with different NBI deposition regions. Moreover, in this study the equilibrium toroidal rotation is assumed to be zero for simplicity. 

\begin{figure}[h!]
\centering
\includegraphics[width=0.45\textwidth]{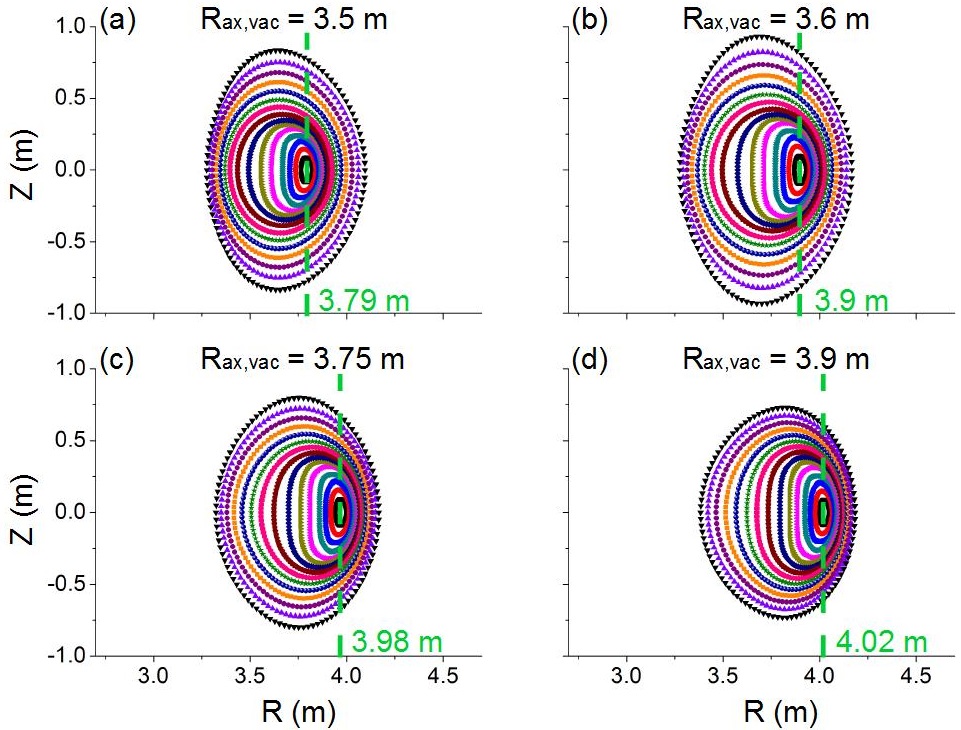}
\caption{Magnetic surfaces of the equilibria at the horizontally elongated cross-section with $R_{ax} = 3.5$ m (a), $R_{ax} = 3.6$ m (b), $R_{ax} = 3.75$ m (c) and $R_{ax} = 3.9$ m (d).}\label{FIG:4}
\end{figure}

Figure~\ref{FIG:4} shows the magnetic surfaces of the equilibria for the different locations of the vacuum magnetic axis location. The large thermal $\beta$ of the equilibria causes a strong Shafranov shift and the deformation of the magnetic surfaces \cite{95}. The magnetic axis is displaced outward with respect to the vacuum location, for example, from $3.5$ to $3.79$ m in the inward shifted case (panel a). It should be noted that the tilt of the tangential NBIs is fixed in LHD, though the NBI deposition region changes in configurations with different vacuum magnetic axis locations, thermal $\beta$ or the beam penetration depth in the plasma core (determined by the plasma density and temperature). 

\subsection{Simulations parameters}

The dynamic toroidal modes ($n$) range from $n=1$ to $4$ and the dynamic poloidal modes ($m$) are selected to cover all the resonant rational surfaces, although the poloidal modes $m=2$ and $3$ of the $n=1$ family and the $m=3$ of the $n=2$ family are always included, even if the respective rational surfaces are non resonant, due to their strong destabilizing effect. The dynamic modes in the simulations indicate the modes of the perturbation that evolve in time while the equilibrium modes describe the equilibria and do not evolve in time. The dynamic mode selection changes between simulations because the NBI current drive modifies the $\rlap{-} \iota$ profile. The equilibrium modes are $n=0$ and $m=[0,6]$ for all the simulations. The simulations are performed with a uniform radial grid of 1000 points. In the following, the mode number notation is $n/m$, which is consistent with the $\iota=n/m$ definition for the associated resonance. 

The closure of the kinetic moment equations (6) and (7) breaks the MHD parities thus both parities must be included for all the dynamic variables. Consequently, the different parities of a mode can show different growth rates and real frequencies in the eigenmode time series analysis. The convention of the code with respect to the Fourier decomposition is, in the case of the pressure eigenfunction, that $n > 0$ corresponds to $cos(m\theta + n\zeta)$ and $n<0$ corresponds to $sin(-m\theta - n\zeta)$. For example, the Fourier component for mode $-1/2$ is $\cos(-1\theta + 2\zeta)$ and for the mode $1/-2$ is $\sin(-1\theta + 2\zeta)$. 

The magnetic Lundquist number ($S$) is a parameter linked to the plasma resistivity (a large $S$ value indicates a plasma with low resistivity). The stability of the PM is affected by the plasma resistivity and the PM growth rate decreases as the S value increases, although the AE growth is weakly affected. The $S$ value in LHD plasma ranges from $10^{8}$ in the core to $10^{6}$ at the periphery. In the simulations $S=5\cdot 10^6$ is assumed, so the PM stability is correctly reproduced for modes located between the middle-outer plasma, although the growth rate of the modes located in the plasma core can be overestimated.

\section{NBI current drive versus plasma stability for an on-axis NBI deposition \label{sec:NBI_onaxis}}

In this section the stability of the PM and AE are analyzed in a case with on-axis NBI injection and different NBCD intensities for inward shifted, default and outward shifted configurations. It should be noted that this study shows hypothetical LHD discharges where the NBI tilt can be modified to maintain an on-axis NBI deposition even if the vacuum magnetic axis changes. That way, the effect of the NBCD on the PM/AE stability can be isolated.

\begin{figure}[h!]
\centering
\includegraphics[width=0.45\textwidth]{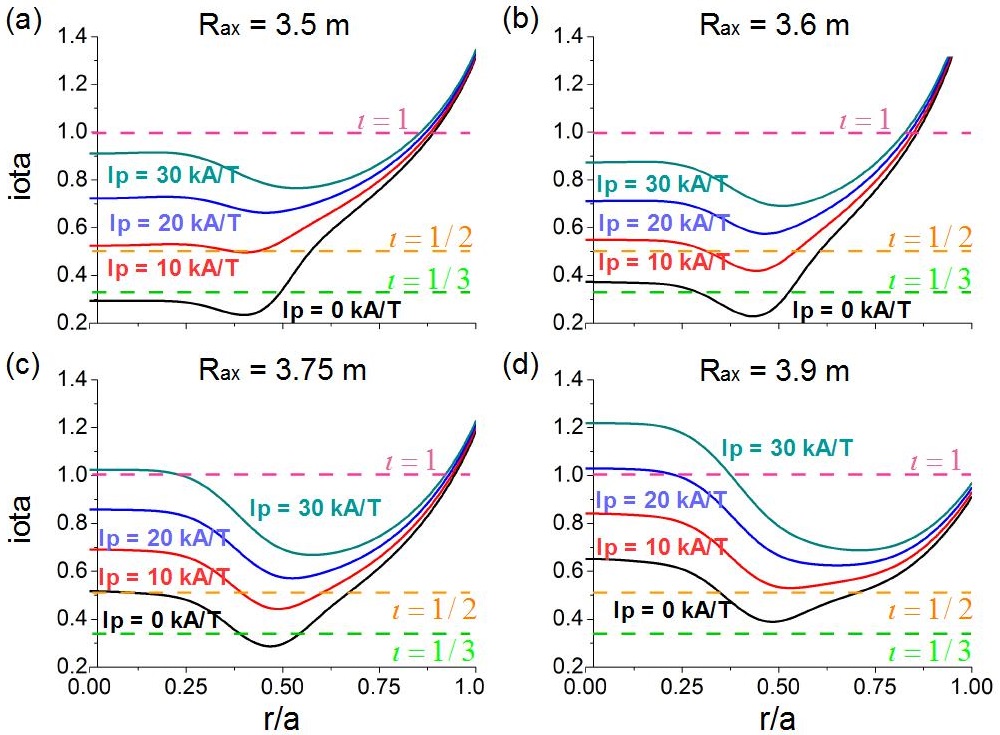}
\caption{Iota profile in inward shifted (a), default (b), outward shifted with $R_{ax}=3.75$ m (c) and outward shifted with $R_{ax}=3.9$ m (d) operation scenarios for different on-axis NBI current drives.}\label{FIG:5}
\end{figure}

\begin{figure*}[h!]
\centering
\includegraphics[width=0.8\textwidth]{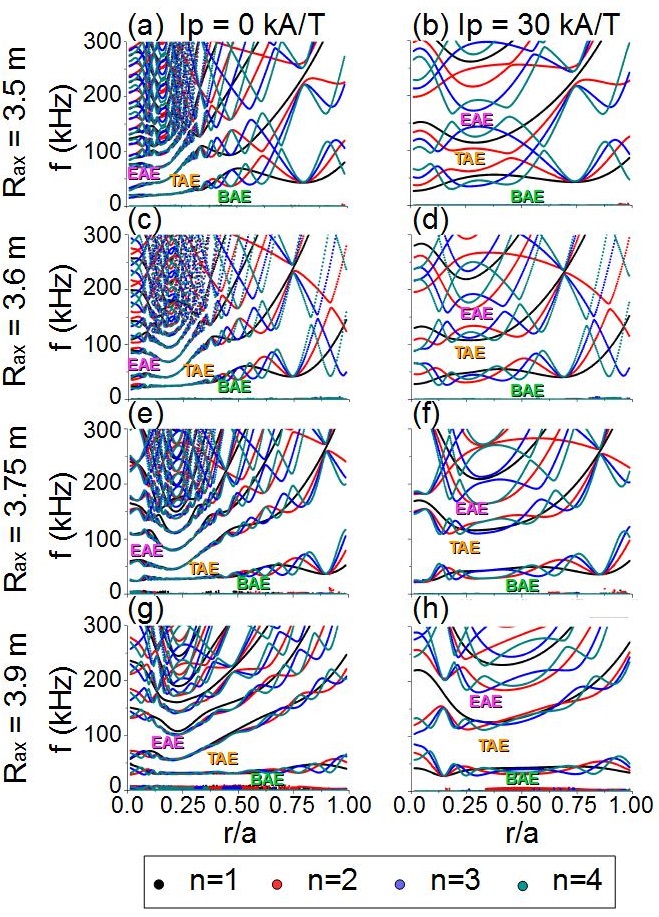}
\caption{Continuum gaps in inward shifted configurations with $I_{p} = 0$ kA/T (a) and $I_{p} = 30$ kA/T (b), default configurations with $I_{p} = 0$ kA/T (c) and $I_{p} = 30$ kA/T (d), outward shifted configurations ($R_{ax}=3.75$ m) with $I_{p} = 0$ kA/T (e) and $I_{p} = 30$ kA/T (f), outward shifted configurations ($R_{ax}=3.9$ m) with $I_{p} = 0$ kA/T (g) and $I_{p} = 30$ kA/T (h).}\label{FIG:6}
\end{figure*}

Figure~\ref{FIG:5} shows the $\rlap{-} \iota$ profiles of the cases analyzed. The rotational transform profile between the middle plasma and the magnetic axis is strongly distorted, leading to an increase of $\rlap{-} \iota_{0}$ in the co-NBCD cases and modifying the magnetic shear. If the magnetic shear increases (decreases), the width of the PM and AE eigenfunctions becomes narrower (broader), improving (deteriorating) the plasma stability. For some cases with a co-NBCD intensity above a given threshold, the $1/2$ rational surface is not present in the plasma thus the destabilizing effect of this mode decreases. In addition, the Alfv\'en gap structure is disturbed; see figure~\ref{FIG:6}, leading to a frequency shift of the Alfv\'en gaps and the consequent variation of the frequency range of the AE families (EP resonances). For example, comparing the cases with balanced NBCD and co-NBCD with $I_{p}=30$ kA/T for an inward shifted configuration (panels a and b), the range of frequencies where an $n=1$ TAE (black lines) can be destabilized in the inner plasma region for the balanced NBCD case goes from $30$ kHz to $50$ kHz, although in the co-NBCD case the frequency range goes from $30$ to $150$ kHz. The general trend shows wider TAE and Elliptical AE (EAE) gaps between the magnetic axis and the middle plasma region as the NBCD increases. In addition, a variation of the frequency range of the Alfv\'en gaps modifies the effect of the continuum damping on the mode stability. If the Alfv\'en gaps are broader, the AEs can be destabilized in frequency ranges where the modes have a wider eigenfunction not intersecting with the continua. On the other hand, if the Afven gaps are slender and there is a frequency shift throughout $r/a$, the AE eigenfunction is narrower and intersections with the continuum can increase damping and avoid the destabilization of the mode. For example, comparing the TAE gap of the balanced NBCD and co-NBCD with $I_{p}=30$ kA/T for the outward shifted configurations (panels e to h), the continuum damping in the co-NBCD case has a weaker effect on the TAE stability. Consequently, the distortion of the $\rlap{-} \iota$ profile caused by the NBCD should affect both the PM and AE stability.

\begin{figure*}[h!]
\centering
\includegraphics[width=0.8\textwidth]{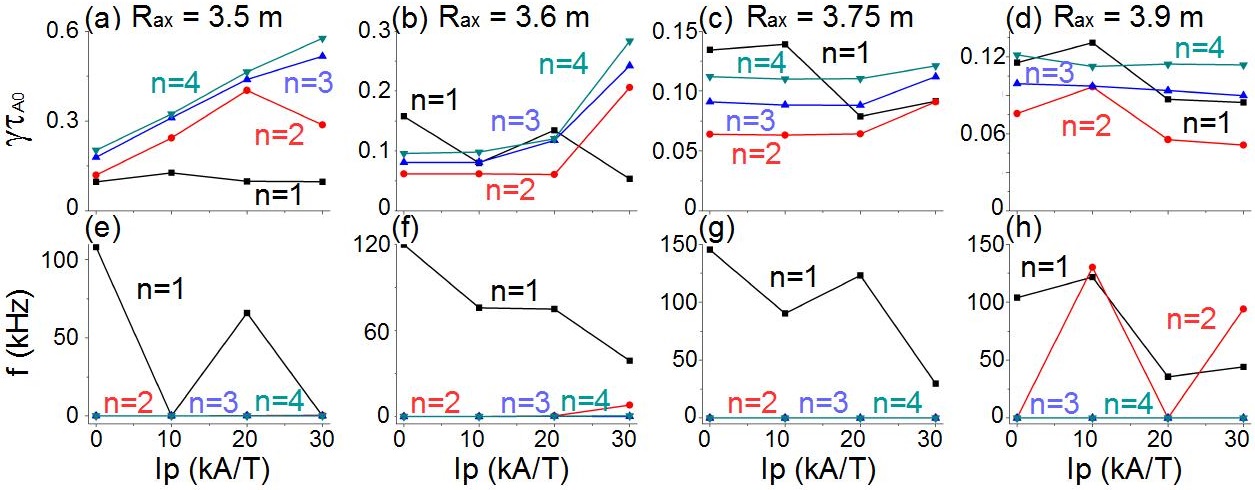}
\caption{Instability growth rate and frequency in inward shifted (panels a and e), default (panels b and f) and outward shifted with $R_{ax}=3.75$ m (panels c and g) and outward shifted with $R_{ax}=3.9$ m (panels d and h) configurations if $\beta_{f0} = 0.01$.}\label{FIG:7}
\end{figure*}

\begin{figure}[h!]
\centering
\includegraphics[width=0.45\textwidth]{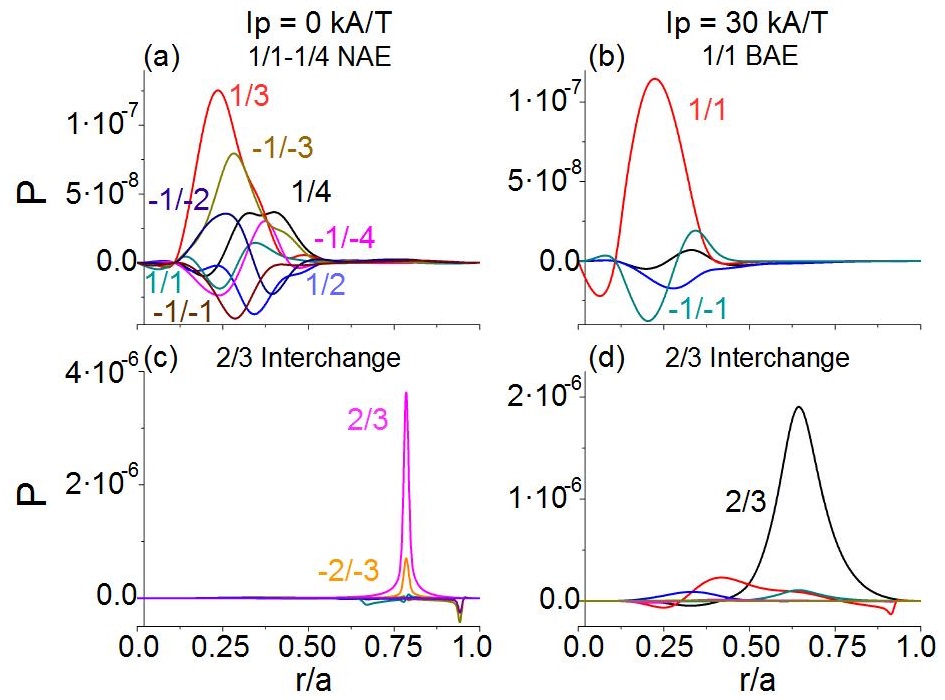}
\caption{Pressure eigenfunction in outward shifted LHD configurations with $R_{ax}=3.75$ m: $n=1$ AE in the balanced (a) and $I_{p}=30$ kA/T (b), $n=2$ interchange mode in the balanced (c) and $I_{p}=30$ kA/T (d)}\label{FIG:8}
\end{figure}

Figure~\ref{FIG:7} indicates the growth rate and frequency of the dominant modes in simulations with different NBCD intensities and locations of the vacuum magnetic axis if $\beta_{f0} = 0.01$, calculated using the initial value solver option of the code. The inward shifted and default LHD configurations show similar trends, although they differ compared to the outward shifted configurations. The $n=2$ to $4$ modes are PM because their frequency is very low compared to the AEs. The $n=2$ to $4$ PM are the dominant modes for all the LHD configurations, except for the $n=2$ AE that is the dominant mode in outward shifted configurations with $R_{ax}=3.9$ m. The growth rate of the $n=2$ to $4$ PM increases with the NBCD intensity caused by the decrease of the magnetic shear between the middle and outer plasma region (see figure~\ref{FIG:5}). Nevertheless, the enhancement of the PM growth rate with the NBCD intensity in the outward shifted cases is weaker with respect to the inward shifted and default configurations. For the $n=1$ toroidal family, the dominant mode is an AE if the NBCD is balanced or $I_{p} =20$ kA/T for the inward shifted and default configurations and for all the NBCD tested in the outward shifted cases. There is an alternation of dominant AEs and PM in inward shifted and default configurations as the intensity of the NBCD increases, caused by the interplay between a weaker magnetic shear and the broadening of the Alfv\'en gaps. It should be noted that the frequency of the $n=1$ AE decreases as the intensity of the NBCD increases, leading to a transition between different AE families, for example in the outward shifted case with $R_{ax} = 3.75$ m where a Non Circular Alfv\'en Eigenmode (NAE, high frequency AE with coupled $m$ and $m+3$ poloidal modes, see fig~\ref{FIG:8}) evolves to a Beta Induced Alfv\'en Eigenmode (BAE, low frequency AE with a single poloidal mode). The transition between dominant modes happens due to the weakening of the continuum damping for the modes in the BAE gap, particularly around $30$ kHz, because the $I_{p} =30$ kA/T case shows a wider BAE gap compared to the NBCD balanced case. Previous studies indicated that the dominant instabilities in strongly outward shifted configurations are high n / helical ballooning modes destabilized at the plasma periphery if the thermal beta is above $0.025$ \cite{96,97,98}. The thermal $\beta$ in the present simulations is above this threshold, although the stability of the high n / helical ballooning modes is weakly affected by the modification of the iota induced by the NBCD at the plasma periphery even if the NBI is deposited off-axis. A topic of a future analysis will be the stabilizing effect of the EP on the high n / helical ballooning modes.

Figure~\ref{FIG:8} shows the pressure eigenfunction of the $n=1$ AE in outward shifted configurations with $R_{ax} = 3.75$ m if the NBCD is balanced (panel a) and $I_{p}=30$ kA/T (panel b), as well as the $n=2$ interchange mode if the NBCD is balanced (panel c) and $I_{p}=30$ kA/T (panel d) cases. The $n=1$ AEs are destabilized in the inner  plasma region because the EP density gradient is located near the magnetic axis. An increase of the NBCD intensity leads to a transition from a $1/1-1/4$ NAE to a $1/1$ BAE. With respect to the $n=2$ toroidal family, a $2/3$ interchange mode is destabilized at the plasma periphery, although the eigenfunction width normalized to the minor radius (defined as $w_{P}/a$) is $0.05$ in the balanced NBCD case, narrower compared to the $I_{p}=30$ kA/T case ($w_{P}/a = 0.25$), located between $r/a =0.5-0.75$. The eigenfunction is broader because the magnetic shear at the plasma periphery decreases as the NBCD intensity increases. The dominant mode of the $n=3$ and $n=4$ families are also PM located at the plasma periphery (data not shown), destabilized by the modes $3/4$ and $4/5$, showing broader eigenfunctions as the NBCD intensity increases and the magnetic shear is weakened.

\begin{figure*}[h!]
\centering
\includegraphics[width=1\textwidth]{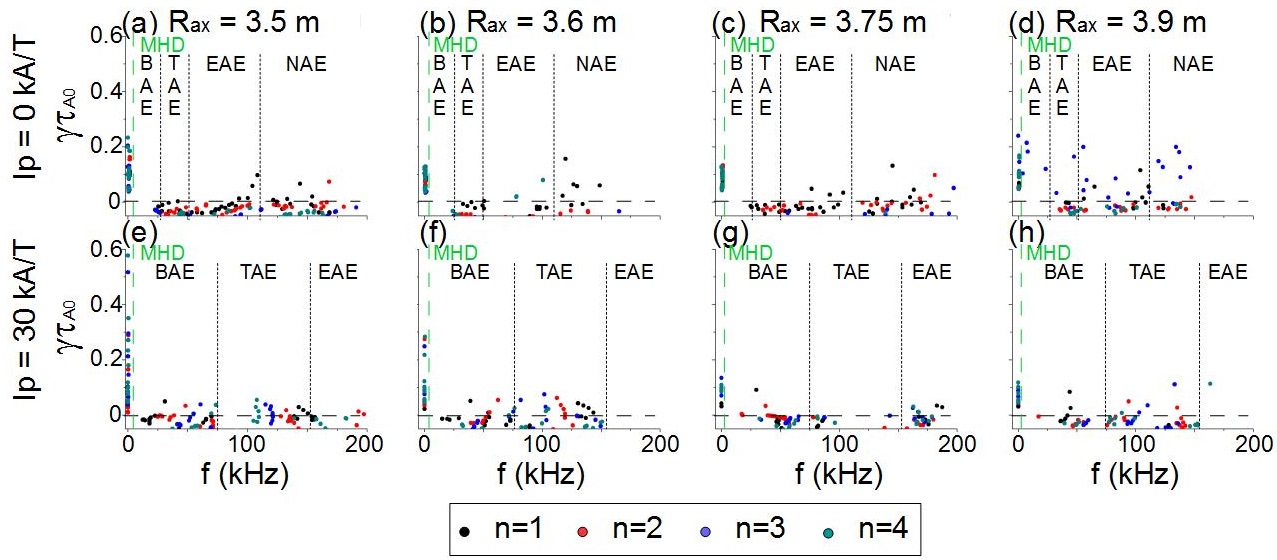}
\caption{Growth rate and frequency of the subdominant modes in inward shifted configurations for a balanced NBCD (a) and a $I_{p}=30$ kA/T (e), default configurations for a balanced NBCD (b) and a $I_{p}=30$ kA/T (f), outward shifted with $R_{ax} = 3.75$ m for a balanced NBCD (c) and a $I_{p}=30$ kA/T (g), outward shifted with $R_{ax} = 3.9$ m for a balanced NBCD (d) and a $I_{p}=30$ kA/T (h). The straight short dashed vertical lines indicate the averaged frequency range of the AE families.}\label{FIG:9}
\end{figure*}

Next, the stability of the subdominant modes is analyzed. These are analyzed by running FAR3d with the eigenvalue solver, instead of the initial value solver. Figure~\ref{FIG:9} indicates the growth rate and frequency of the subdominant modes in the inward shifted, default, outward shifted ($R_{ax} = 3.75$ m and $R_{ax} = 3.9$ m) configurations for a balanced NBCD and co-NBCD of $I_{p}=30$ kA/T. The straight short dashed vertical lines indicate the averaged frequency range of the AE families (the frequency range of the AE families has a radial dependency so these lines are for indicative purposes only). An increase of the NBCD intensity up to $I_{p}=30$ kA/T leads to AEs with an averaged lower growth rate with respect to the balanced NBCD case. The growth rate of the high frequency AEs (EAEs and NAEs) decreases (EAEs can be identified by the coupling between $m$ and $m+2$ poloidal mode), although low frequency AEs (BAE) and TAEs are destabilized. Hence, the NBCD improves the AE stability of the $n=1$ and $n=2$ AEs although higher $n$ AEs are destabilized. Nevertheless, the outward shifted configurations show a more robust improvement of the AE stability as the NBCD intensity increases because the AEs growth rate is smaller with respect to the balanced NBCD case.

\begin{figure}[h!]
\centering
\includegraphics[width=0.45\textwidth]{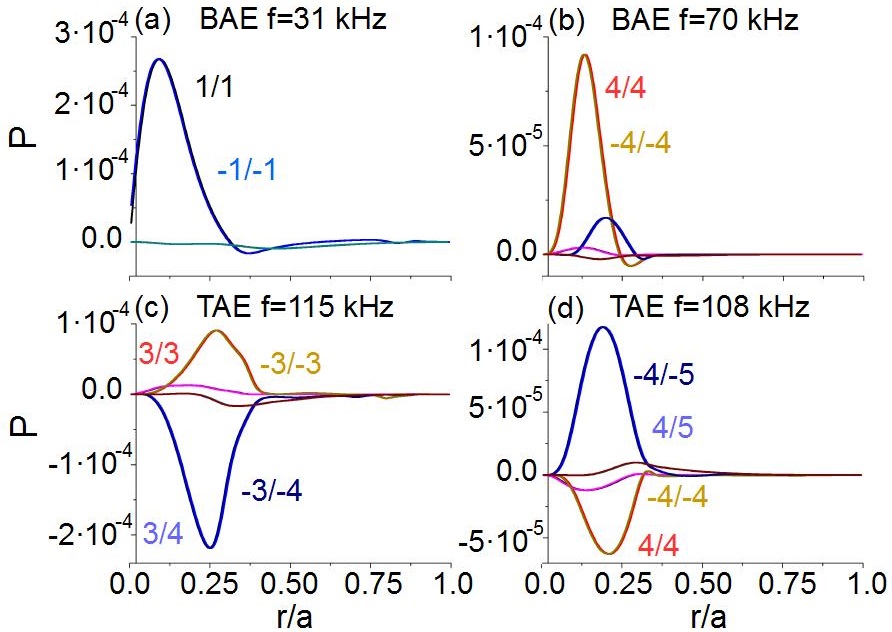}
\caption{Pressure eigenfunction of some subdominant AEs in the inward shifted LHD configuration if $I_{p}=30$ kA/T (a) $1/1$ BAE (b), $4/4$ BAE, (c) $3/3 - 3/4$ TAE and (d) $4/4 - 4/5$ TAE.}\label{FIG:10}
\end{figure}

Figure~\ref{FIG:10} shows the pressure eigenfunction of some subdominant AE destabilized in the inward shifted configuration for an NBCD of $I_{p}=30$ kA/T. The AE are destabilized in the inner region of the plasma. The $n=1$ to $n=4$ BAE are overtones of the same instability, linked to the $\rlap{-} \iota = 1$ rational surface. The $3/3-3/4$ and $4/4-4/5$ TAEs are destabilized slightly outward with respect to the BAEs.

\begin{figure*}[h!]
\centering
\includegraphics[width=0.95\textwidth]{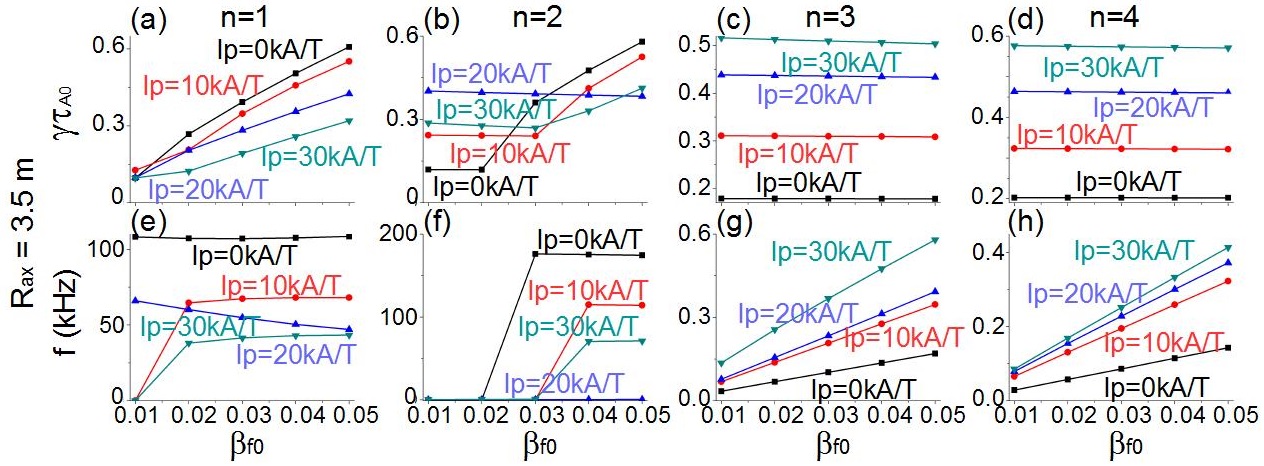}
\caption{Instability growth rate and frequency in inward shifted configurations for different NBCD intensities and EP $\beta$.}\label{FIG:11}
\end{figure*}

\begin{figure*}[h!]
\centering
\includegraphics[width=0.95\textwidth]{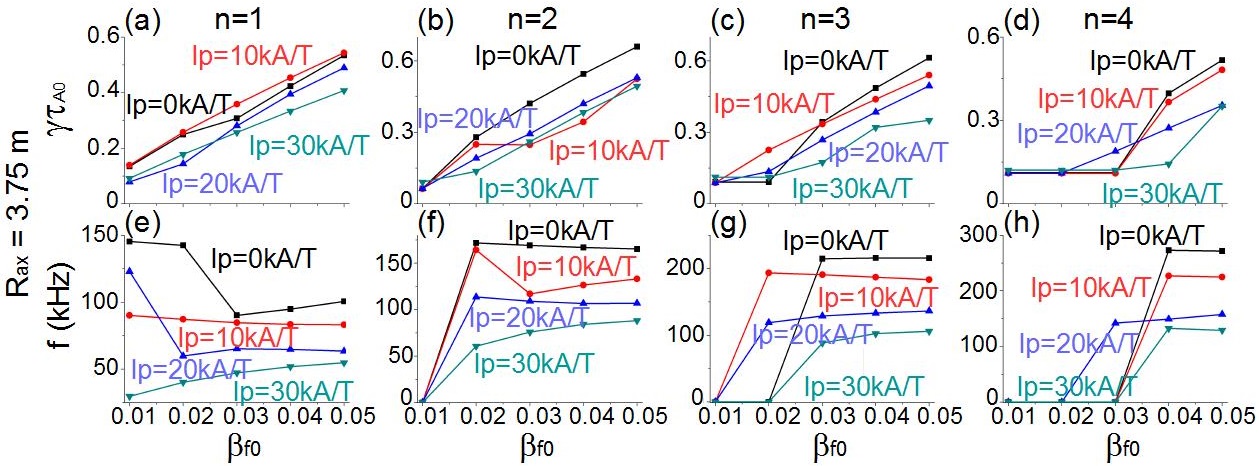}
\caption{Instability growth rate and frequency in the outward shifted configurations $R_{ax} = 3.75$ m for different NBCD intensities and EP $\beta$.}\label{FIG:12}
\end{figure*}

\begin{figure*}[h!]
\centering
\includegraphics[width=0.95\textwidth]{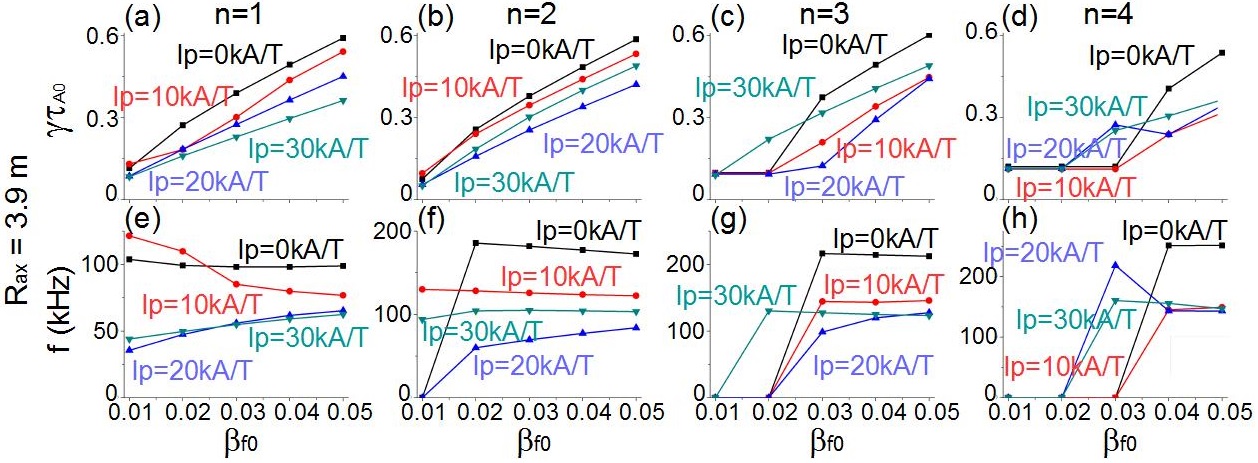}
\caption{Instability growth rate and frequency in the outward shifted configurations $R_{ax} = 3.9$ m for different NBCD intensities and EP $\beta$.}\label{FIG:13}
\end{figure*}

The EP $\beta$ is linked to the NBI injection intensity, thus a higher EP $\beta$ means a stronger EP destabilizing effect. It is important to identify the EP $\beta$ threshold to destabilize the AEs in the configurations analyzed. Figures~\ref{FIG:11},~\ref{FIG:12} and~\ref{FIG:13} show the growth rate and frequency of the $n=1$ to $n=4$ instabilities in inward and outward shifted configurations for different NBCD intensities and EP $\beta$ values as the EP density increases (these plots include the $\beta_{f0} = 0.01$ data of fig~\ref{FIG:7}). In the inward shifted configuration, the $n=1$ toroidal family is AE unstable for all the NBCD intensities analyzed from an EP $\beta$ of $0.02$, although only for the balanced NBCD case the $n=1$ EAE is unstable, changing to an $n=1$ TAE if $I_{p}>0$ kA/T. Once the AE is triggered, the mode growth rate increases with the EP $\beta$ because the population of EP destabilizing the AE is larger compared to simulations with lower EP $\beta$. It should be noted that the AE growth rate decreases as the NBCD intensity increases. For the $n=2$ toroidal family, an NAE is destabilized in the balanced NBCD case from a $\beta_{f0} = 0.03$, EAEs if $I_{p}=10$ and $30$ kA/T cases from an EP $\beta = 0.04$, although in the configuration with $I_{p}=20$ kA/T the $n=2$ toroidal family is AE stable at least up to a $\beta_{f0} = 0.05$. In the same way, the $n=3$ and $n=4$ toroidal families are AE stable for all the NBCD intensities and EP $\beta$ analyzed (the mode frequency is below $1$ kHz, see panels g and h). The growth rates that are almost independent of the EP $\beta$ value imply the mode is a PM. Regarding the outward shifted configurations, the $n=2$ toroidal family is AE unstable for all the NBCD intensities analyzed from a EP $\beta$ of $0.02$ ($0.01$ if the NBCD is $I_{p}=10$ or $30$ kA/T in the case with $R_{ax} = 3.9$ m). In addition, the $n=3$ and $n=4$ toroidal families are AE unstable for all the NBCD intensities tested from $\beta_{f0} = 0.03$ and $0.04$, respectively. Thus, the outward shifted configurations show a lower threshold of the EP $\beta$ for the destabilization of AE compared to the inward shifted configurations (the analysis of the EP $\beta$ threshold for the default configuration is not included because the trends are similar to the inward shitted configuration). The lower threshold of the EP $\beta$ can be explained by a decrease of the continuum damping as the NBCD intensity increases (broader Alfv\'en gaps in the inner-middle plasma region, see fig~\ref{FIG:6}e to h).

In summary, a larger intensity of the NBCD in the inward shifted and default configuration leads to an enhancement of the PM (fig~\ref{FIG:7}), although the AEs growth rate is lower (fig~\ref{FIG:9}). In addition, the threshold to destabilize the $n=1$ and $2$ AEs with respect to the EP $\beta$ is higher as the NBCD intensity increases in the inward shifted configuration (fig~\ref{FIG:11}). Outward shifted configurations show a lower EP $\beta$ threshold as the NBCD intensity increases, see fig~\ref{FIG:12} and~\ref{FIG:13}. On the other hand, the PM and AE growth rate decreases as the NBCD intensity increases (fig~\ref{FIG:9}). It should be noted that the analysis is not fully self-consistent since an increase of the NBCD intensity should be associated with a higher beam power and larger EP $\beta$, reason why the study of the EP $\beta$ threshold is also performed.

\section{NBI current drive versus plasma stability for an off-axis NBI deposition \label{sec:NBI_offaxis}}

The LHD discharges with a high $\beta$ or an inward shifted vacuum magnetic axis location lead to off-axis NBI depositions. Consequently, the effect of the off-axis NBI deposition must be included in the analysis. In this section the stability of the PM and AE is analyzed with respect to the NBCD if the NBI is injected off-axis.

\begin{figure}[h!]
\centering
\includegraphics[width=0.45\textwidth]{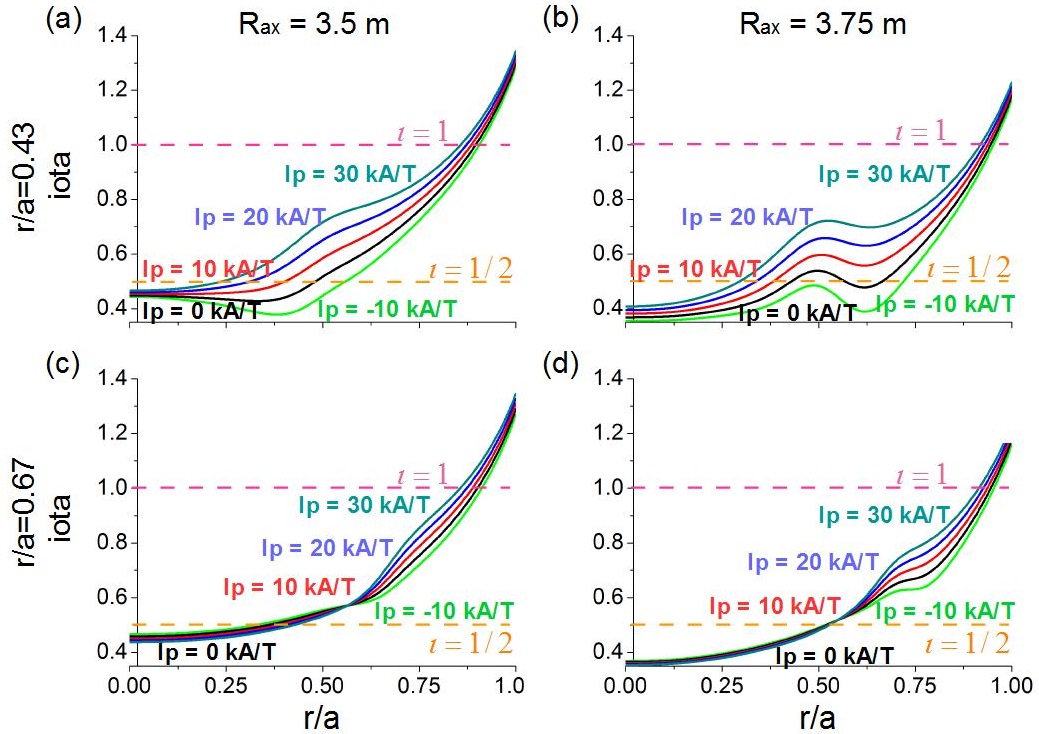}
\caption{Iota profile of the inward shifted configuration if the NBI is deposited at $r/a = 0.43$ (a) or $0.67$ (c). Iota profile of the outward shifted configuration with $R_{ax} = 3.75$ m if the NBI is deposited at $r/a = 0.43$ (b) or $0.67$ (d).}\label{FIG:14}
\end{figure}

\begin{figure*}[h!]
\centering
\includegraphics[width=0.8\textwidth]{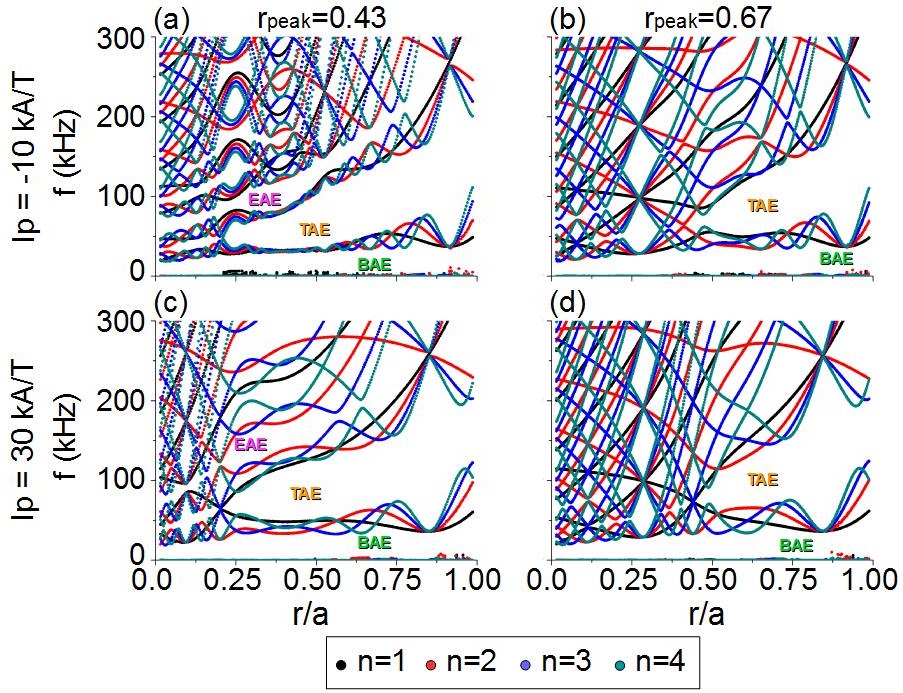}
\caption{Continuum gaps of the outward shifted configurations with $R_{ax} = 3.75$ m if the NBCD is $I_{p} = -10$ kA/T and the NBI is deposited at $r_{peak} = 0.43$ (a), at $r_{peak} = 0.67$ (b), or the NBCD is $I_{p} = 30$ kA/T and the NBI is deposited at $r_{peak} = 0.43$ (c), at $r_{peak} = 0.67$ (d).}\label{FIG:15}
\end{figure*}

Figure~\ref{FIG:14} shows the $\rlap{-} \iota$ profile of inward (panels a and c) and outward shifted configurations with $R_{ax} = 3.75$ m (panels b and d) if the NBI is deposited off-axis at $r/a = 0.43$ and $0.67$ (these $r_{peak}$ values are selected representing NBI depositions in the middle and outer plasma region). The distortion of the $\rlap{-} \iota$ profile is weaker compared to an on-axis NBI injection and the $1/2$ rational surface is present in all the cases. The effect of the NBCD is smaller as the beam is deposited further outwards because the rotational transform is more strongly determined by the coils. Figure~\ref{FIG:15} indicates the continuum gaps of an outward shifted configuration with $R_{ax} = 3.75$ m if the NBI is deposited at $r/a = 0.43$ and $0.67$ for a ctr-NBCD with $I_{p} = -10$ kA/T (panels a and b) and a $I_{p} = 30$ kA/T (panels c and d). The Alfv\'en gaps are slender between the inner and the middle plasma as the NBI is deposited further away from the magnetic axis, leading to a large enhancement of the continuum damping at the core region. On the other hand, the increase of the co-NBCD intensity leads to wider Alfv\'en gaps (same trend compared to the on-axis cases, see fig.~\ref{FIG:6}), although the effect is weaker as the NBI is injected closer to the plasma periphery, thus the stability of the AE/EPM is weakly affect by the NBCD intensity by means of the variation of the EP resonance and continuum damping. It should be noted that the continuum gap structure is strongly modified by a change of the NBI deposition region compared to a change of the NBCD intensity, although these trends must be analyzed separately.

\begin{figure*}[h!]
\centering
\includegraphics[width=0.95\textwidth]{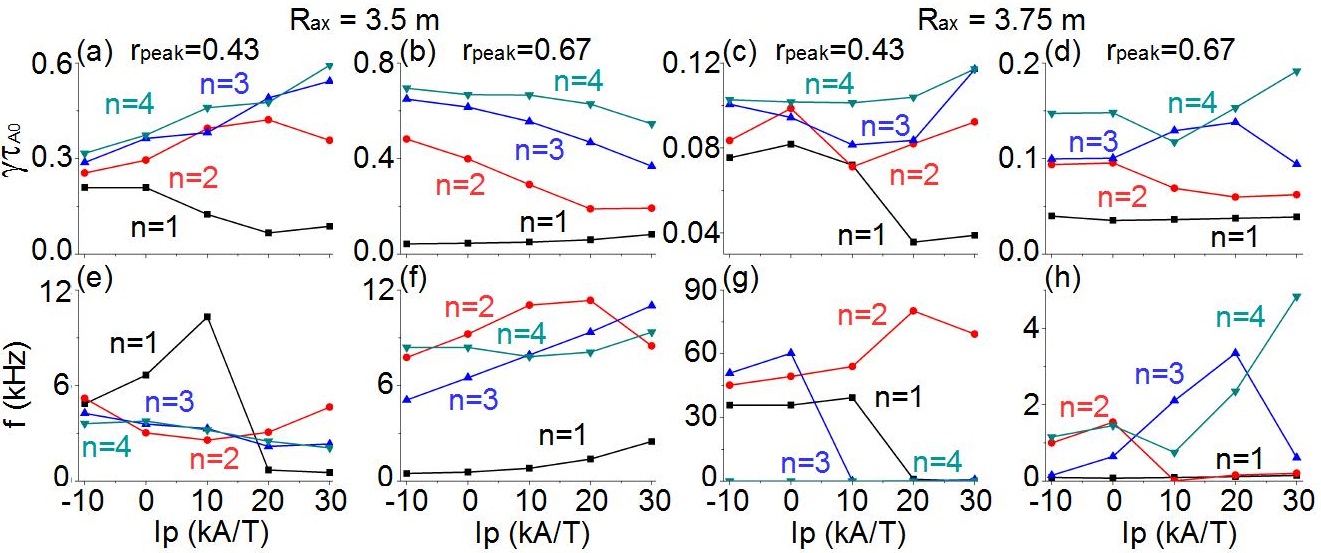}
\caption{Instability growth rate and frequency in the inward shifted configuration with the NBI deposited in the middle plasma (panels a and e) and at the plasma periphery (panels b and f). Instability growth rate and frequency in the outward shifted configuration $R_{ax} = 3.75$ m and the NBI deposited at the middle plasma (panels c and g) and at the plasma periphery (panels d and h).}\label{FIG:16}
\end{figure*}

Figure~\ref{FIG:16} shows the growth rate and frequency of the dominant modes in simulations with different NBCD intensities, locations of the vacuum magnetic axis and NBI deposition regions. For the inward shifted configuration with the NBI deposited at the middle plasma, the increase of the co-NBCD intensity leads to an enhancement of the $n=2$ to $n=4$ AE/EPMs, because these modes are destabilized between $r/a=0.5$ and $0.75$ where the magnetic shear decreases. On the other hand, the $n=1$ AE/EPM is stabilized if the $I_{p} \ge 20$ kA/T because this mode is localized between $r/a=0.4$ and $0.5$ where the magnetic shear increases. If the NBI is deposited at the outer plasma, the growth rate of the $n=2$ to $n=4$ AE/EPMs decreases as the co-NBCD intensity increases due to an enhanced magnetic shear around $r/a=0.65$, although the $n=1$ AE is destabilized if the $I_{p} \ge 20$ kA/T because the magnetic shear at the plasma periphery decreases. In the outward shifted configuration with $R_{ax} = 3.75$ m and the NBI deposited at the middle plasma, there is a transition between dominant $n=1$ and $n=3$ AEs to dominant PM as the co-NBCD intensity increases, although the $n=2$ AE is still dominant and the $n=4$ PM growth rate slightly increases. If the NBI is deposited at the outer plasma, the $n=2$ to $n=3$ AE/EPMs are stabilized as the co-NBCD intensity increases although the growth rate of the $n=4$ AE/EPM increases.

\begin{figure}[h!]
\centering
\includegraphics[width=0.45\textwidth]{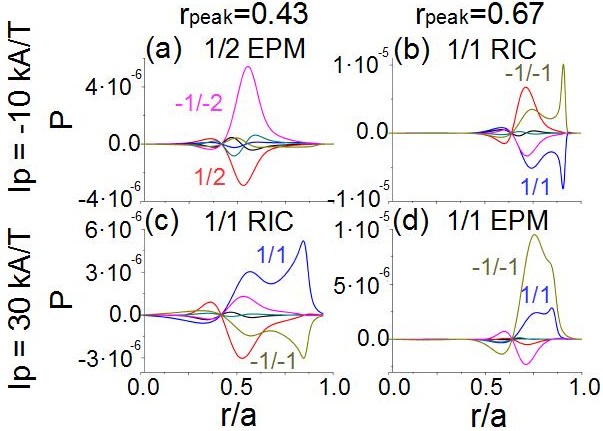}
\caption{Pressure eigenfunction of the $n=1$ instability in the inward shifted configurations ($R_{ax} = 3.5$ m) if the NBCD $I_{p}=-10$ kA/T for a NBI deposited at the middle plasma (a) or at the plasma periphery (b), if the NBCD $I_{p}=30$ kA/T for a NBI deposited at the middle plasma (c) or at the plasma periphery (d).}\label{FIG:17}
\end{figure}

Figure~\ref{FIG:17} shows the pressure eigenfunction of the $n=1$ instability in the inward shifted configuration ($R_{ax} = 3.5$ m) if the NBI is deposited at the middle or the outer plasma region and the NBCD intensity is $I_{p}=-10$ or $30$ kA/T. The simulations with the NBI deposited at the middle plasma show a $1/2$ EPM located at the middle plasma if $I_{p}=-10$ kA/T (panel a), although a $1/1$ resistive interchange modes (RIC) is triggered if $I_{p}=30$ kA/T (panel c). On the other hand, the simulations with the NBI deposited at the outer plasma show an unstable $1/1$ RIC if $I_{p}=-10$ kA/T (panel b) although a $1/1$ EPM located at the plasma periphery is triggered if $I_{p}=30$ kA/T (panel d). Such $1/2$ and $1/1$ AEs are in the same frequency range with respect to the $1/2$ and $1/1$ EIC observed in LHD discharges. The EIC are destabilized by helically trapped particles generated by the perpendicular beam in LHD discharges \cite{43,92}, although in the present model the resonance is caused by the passing EP generated by the tangential beam. These results suggest the possibility that the EIC can also be destabilized by passing EP.

\begin{figure}[h!]
\centering
\includegraphics[width=0.35\textwidth]{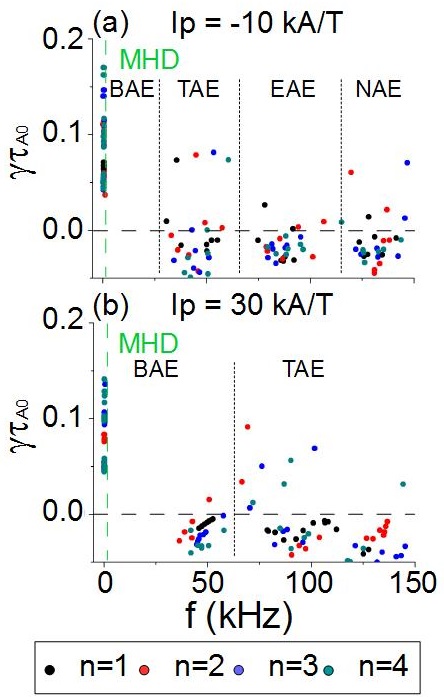}
\caption{Growth rate and frequency of the subdominant modes in the outward shifted configurations with $R_{ax} = 3.75$ m for a ctr-NBCD with $I_{p}=-10$ kA/T (a) or a co-NBCD with $I_{p}=30$ kA/T (b) if the NBI is deposited in the middle plasma region ($r_{peak} = 0.43$). The straight short dashed vertical lines indicate the averaged frequency range of the AE families.}\label{FIG:18}
\end{figure}

\begin{figure}[h!]
\centering
\includegraphics[width=0.45\textwidth]{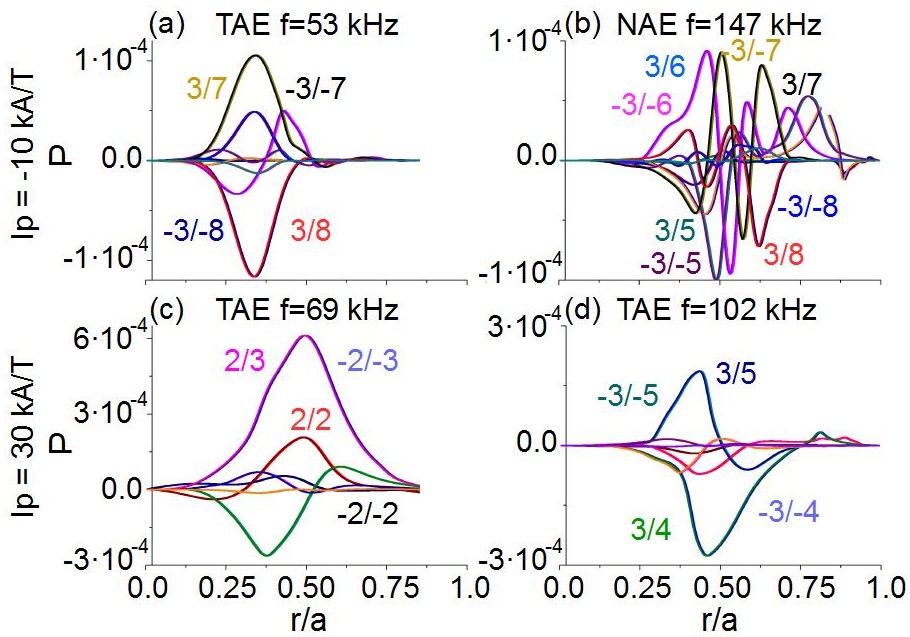}
\caption{Pressure eigenfunction of the subdominant AEs in the outward shifted configuration $R_{ax} = 3.75$ m for a ctr-NBCD deposited in the middle plasma region ($r_{peak} = 0.43$) with $I_{p}=-10$ kA/T (a) $3/7-3/8$ TAE and (b) $3/5-3/8$ NAE. For a co-NBCD deposited in the middle plasma region ($r_{peak} = 0.43$) with $I_{p}=30$ kA/T (c) $2/2-2/3$ TAE and (d) $3/4-3/5$ TAE.}\label{FIG:19}
\end{figure}

Figure~\ref{FIG:18} shows the subdominant modes in the outward shifted configuration with $R_{ax} = 3.75$ m and the beam deposited at the middle plasma region ($r_{peak} = 0.43$) for a ctr-NBCD of $I_{p}=-10$ kA/T (panel a) and a co-NBCD of $I_{p}=30$ kA/T (panel b). The analysis indicates a partial improvement of the AE stability as the co-NBCD intensity increases, because the simulations with a co-NBCD of $I_{p}=30$ kA/T show stable $n=1$ AEs and only the $n=2$ to $n=4$ TAEs are unstable (except a marginally unstable $n=2$ BAE). In addition, there is a decrease of the growth rate of the $n=3$ and $n=4$ TAEs ($10 \%$ and $30 \%$, respectively), although the growth rate of the  $n=2$ TAE is almost the same). Compared to the simulations with an on-axis NBI deposition and a co-NBCD of $I_{p}=30$ kA/T (see fig~\ref{FIG:9}d), the $n=1$ AEs are stable although the $n=2$ to $n=4$ AEs are further destabilized. In addition, the low frequency AE (BAE) are stable and the unstable AEs are $n=2$ to $n=4$ TAEs. Consequently, no evident optimization trend is observed in the simulations if the NBI is deposited at the middle plasma with respect to the on-axis cases. Figure~\ref{FIG:19} shows the pressure eigenfunction of the subdominant AEs with the largest growth rates destabilized in the ctr- and co-NBCD cases if the NBI is deposited at $r_{peak}=0.43$. For a ctr-NBCD of $I_{p}=-10$ kA/T, a $3/7-3/8$ TAE with $53$ kHz is unstable in the inner plasma region (panel a) and a $3/5-3/8$ NAE with 147 kHz in the middle plasma region (panel b). For a co-NBCD of $I_{p}=30$ kA/T, a $2/2-2/3$ TAE with $69$ kHz is unstable in the middle plasma region (panel c) and a $3/4-3/5$ TAE with $104$ kHz also in the middle plasma region (panel d).

In summary, if the PM and AE stability is compared between cases with off-axis and on-axis NBI depositions, the general trend indicates that the growth rate of the PM decreases while the AE/EPM growth rate increases as the NBI deposition region is moved out, leading to the destabilization of AE/EPM with a growth rate similar or even larger than the unstable PM in the on-axis case (particularly in the inward shifted model). Consequently, no optimization trend was identified.

\section{Simulation versus experimental trends \label{sec:exp}}

In this section the trends observed in LHD experiments during the co- and ctr-NBCD phases, see fig.~\ref{FIG:2}, are compared with the simulation results. It should be noted that the LHD discharges shown have an average total $\beta$ smaller with respect to the simulations. The difference is caused by a larger EP $\beta$, selected to easily identify the AE stability trends. The simulations averaged total $\beta$ is $1.8 \%$ (table~\ref{Table:1}), while the local maxima of the averaged total $\beta$ during the discharges reaches $0.5 \%$ for the shot $147288$ and $0.25 \%$ for the shot $147372$ (fig~\ref{FIG:2}a). Consequently, the effect of the off-axis NBI deposition due to the Shafranov shift as well as the EP forcing on the PM/AE stability is smaller. Nevertheless, the effect of the NBCD is large enough to cause clear differences between co- and ctr-NBCD phases on the plasma stability.

During the discharge $147288$ there is a ctr-NBI phase between $3.5$ to $5.5$ s leading a ctr-NBCD up to $-20$ kA/T (panel b). The co-NBI injection is dominant after $t = 6$ s generating a co-NBCD up to $20$ kA/T. During the co-NBCD phase the averaged $\beta$ is $0.25 \%$ increasing up to $0.35 \%$ (panel a). The ctr-NBCD phase shows the destabilization of several AEs in the range of frequencies between $30$ and $180$ kHz that correspond with BAE, TAE and EAE in the simulations (see fig~\ref{FIG:6}c), although the AE activity is weaker in the co-NBCD phase where the AE frequency range is displaced to higher values (see fig~\ref{FIG:6}d). The decrease of the magnetic probe signal during the co-NBCD phase is reproduced by the simulations  (see fig~\ref{FIG:9}b and f), showing that the AE growth rate in the balanced current case is larger compared to the co-NBCD case, thus the AEs must be further destabilized in a ctr-NBCD case. Also, the strongest magnetic probe signal in the co-NBCD phase is observed at the frequency range of the BAE (around $75$ kHz), also consistent with the simulations showing the growth rate of the BAE as the largest between the AE families. With respect to the shot $147372$, there is a co-NBCD phase until $t = 6.5$ s with a maximum of $20$ kA/T at $t = 5.5$ s and a ctr-NBCD phase from $t = 6.5$ s with a local maxima of $-10$ kA/T at $t = 7.5$ s. During the co-NBCD phase the averaged $\beta$ is $0.1 \%$ increasing up to $0.25 \%$ during the ctr-NBCD phase. There is a down-shift of the observed frequencies of the different AE families between the co-NBCD to the ctr-NBCD phases, also reproduced in the simulations (see fig~\ref{FIG:6}g and h). The spectrogram of the magnetic probe signal indicates the destabilization of several AEs at frequencies of about $50$, $125$ and $175$ kHz, as well as the further destabilization of high frequency AE during the ctr-NBCD phase, both trends also identified in the simulations (see fig~\ref{FIG:9}d and h). In addition, the stronger spectrogram signal in the discharge $147372$ with respect to the shot $147288$ can be explained by the off-axis NBI injection caused by the outward shifted location of the vacuum magnetic axis, leading to an enhancement of the AE activity reproduced by the simulations as an increase of the AEs growth rate (please compare fig~\ref{FIG:9}g and fig~\ref{FIG:18}b). It should be noted that the modes identified in the simulations as EAE and NAE must be verified experimentally. No clear evidence of the destabilization of EAE and NAE have been found in LHD discharges yet \cite{99}.

The effect of the toroidal current on the PM stability in inward shifted LHD discharges with high thermal $\beta$ was analyzed in previous studies \cite{100,101}, showing the stabilization of the $1/2$ mode and the destabilization of the $2/3$ and $3/2$ modes if the toroidal current increases above $30$ kA/T during the co-NBCD phase. The experimental observations are consistent with the simulation results, showing that the $1/2$ mode stabilization is caused by the increase of the iota profile leading to a non resonant $1/2$ rational surface (see fig~\ref{FIG:5}a and b). In addition, the PM located at the plasma periphery are further destabilized due to a decrease of the magnetic shear, leading to wider instability eigen-functions and extending the unstable plasma region from the $\rlap{-} \iota = 1$ rational surface to the $2/3$ and $3/2$ as the co-NBCD intensity increases. Another study identified the triggering of a minor collapse linked to the $\rlap{-} \iota = 1$ rational surface if the thermal $\beta$ and co-NBCD intensity increases above a given value \cite{50}. Nevertheless, due to the limitations of the present study, the effect of the ctr-NBCD on the PM stability was not analyzed, although experimental observations indicated the enhancement of the $1/2$ mode if the ctr-NBCD intensity increases above $-20$ kA/T for an operational magnetic field of $2.75$ T \cite{102}.

In summary, the diagnostics data obtained during the inward and outward shifted discharges show a reasonable agreement with the trends obtained from the simulations for the AE stability. 

\section{Discussion \label{sec:conclusions}}

The cases with on-axis NBI deposition show the destabilization of $n=2$ to $4$ PM in the middle-outer plasma region, whose growth rate increases as the NBCD intensity is enhanced, particularly in the inward shifted configurations. The $n=1$ toroidal family is AE unstable, especially in the outward shifted configurations, although the growth rate and frequency decrease as the co-NBCD intensity is enhanced, leading to the transition between different AE families. The study of the subdominant modes indicates a lower growth rate of $n=1$ and $2$ AEs as the co-NBCD intensity increases although the $n=3$ and $4$ AEs are destabilized. Also, the cases with a large co-NBCD intensity show a higher threshold of the EP $\beta$ to destabilize AEs and, once destabilized, the AEs growth rate is lower as the co-NBCD intensity increases. It should be noted that the AE stability is worse in outward shifted configurations, destabilized for a lower EP $\beta$ threshold, and all the toroidal families are AE unstable while in the inward shifted configurations only the dominant $n=1$ and $n=2$ instabilities are AEs. This is caused by a weaker stabilizing effect of the continuum damping due to the expansion of the frequency range of the Alfv\'en gaps in the inner-middle plasma region as the NBCD intensity increases.

The analysis of cases where the NBI is deposited off-axis indicates a decrease of the PM growth rate compared to the on-axis cases. On the other hand, the AE/EPMs are further destabilized showing a growth rate similar or even larger than the PM in the on-axis cases. Also, as the NBI is deposited closer to the plasma periphery, the low frequency AE/EPM are further destabilized. The simulations for inward shifted configurations show unstable $n=1$ EPMs in the same frequency range as the $1/1$ and $1/2$ EICs observed in LHD discharges, destabilized by passing EP instead of helically trapped EP. If the beam is deposited in the middle plasma region, a $1/2$ EPM can be stabilized if the co-NBCD intensity is larger than $20$ kA/T, although if the beam is deposited in the plasma periphery a $1/1$ EPM is destabilized if the co-NBCD intensity is larger than $10$ kA/T. It should be noted that the EPMs are stable in the outward shifted case for all the NBCD intensities analyzed. 

The optimization trends of the AE stability with respect to the NBCD intensity identified by the simulations are consistent with the diagnostics data in the inward and outward shifted discharges analyzed. In addition, previous experimental studies dedicated to analyze the PM stability for different NBI co- and ctr-NBCD show consistent results with the simulations. It should be noted that the plasma stability in ctr-NBCD cases was not explored in the study due to the model limitations for reproducing LHD operation scenario with a $\rlap{-} \iota < 0.2$, thus the ctr-NBCD phase of the discharges is not studied in this analysis. Nevertheless, the extrapolation of the plasma stability properties of the balanced current case is consistent with previous studies up to a ctr-NBCD intensity of $-60$ kA/T. The verification of the stability trends identified by the numerical modeling requires the analysis of the PM and AE activity with respect to the ctr- and co-NBCD intensity in dedicated experiments. This study will be the topic of a following communication.

\section{Conclusions and outlook}

A set of linear simulations are performed by the FAR3d code to study the effect of the NBI current drive on the stability of pressure gradient driven modes and Alfv\'en Eigenmodes. Simulation results and experimental data are compared, selecting two LHD shots that represent inward and an outward shifted LHD configurations where the NBCD is not balanced and the plasma current increases during the discharge.

The NBCD strongly modifies the $\rlap{-} \iota$ profile, particularly if the beam is deposited on-axis or in the middle plasma region and the intensity of the driven current is large. The distortion of the rotational transform causes the modification of the magnetic shear and the Alfv\'en gap structure, leading to a non resonant $\rlap{-} \iota = 1/2$ rational surface if the NBI is deposited on axis and the co-NBCD is above a given intensity. Consequently, the simulations show a change of the plasma stability with respect to the PM and AE.

The NBCD is a useful tool in LHD discharges to change the magnetic field topology, leading to the modification of the stability of the PM and AEs. Nevertheless, no optimization trends are identified for both instabilities at the same time, because a larger NBCD intensity leads to the further destabilization of the PM and weaker AEs in inward shifted configurations, while the stability of the PM is almost unchanged while the AEs are enhanced in the outward shifted configurations. Consequently, an optimized operation for inward shifted configurations requires a ctr-NBCD, improving the PM stability, while keeping the ctr-NBCD intensity low enough to avoid the destabilization of strong $n=1$ and $n=2$ AEs. On the other hand, an optimized operation for outward shifted configurations requires a co-NBCD, because the PM stability is weakly affected up to a co-NBCD intensity of $I_{p} = 30$ kA/T while the AE stability improves and the EP $\beta$ threshold increases.

The modification of the rotational transform caused by the NBCD allows steady-state operational scenarios with an improved PM and AE stability, not accessible using the standard magnetic field configuration generated from the coils. Stellarators do not need a net current for steady state operations, although for profile optimization purposes, an alternative beam injection can sustain a net current in the plasma. In addition, the NBCD can be combined with other non inductive currents generated by ECCD or LH for a further optimization of the plasma stability. It should be noted that the effect of the NBCD depends on the magnetic field intensity, because a larger amount of plasma current is required to modify the iota profile as the magnetic field intensity increases. Consequently, LHD operations with a high magnetic field require longer pulses and a strongly unbalanced co- or ctr-NBI injection to generate a net plasma current large enough to affect the PM and AE stability. Nevertheless, if an LHD operation scenario with improved plasma stability is identified for a given plasma current generated by the NBCD, if this amount of plasma current is held approximately constant during the flat phase of the discharge (for example using a NBI pattern with cycles of dominant co- and ctr- NBI injection), the $\beta$ limit of the operation scenario can be improved compared to a similar discharge using NBI patterns that lead to a balanced NBCD. This optimization strategy depends on the pulse length and the amount of plasma current driven by the NBCD with respect to the magnetic field intensity, so the application of this method is more demanding in fusion devices with large magnetic fields or a Tokamak with large plasma currents. Dedicated experiments will be performed in future LHD campaigns to analyze in further detail the optimization trends suggested in the present study.

\section*{Appendix}

\subsection*{Mode identification}

\begin{figure*}[h!]
\centering
\includegraphics[width=0.95\textwidth]{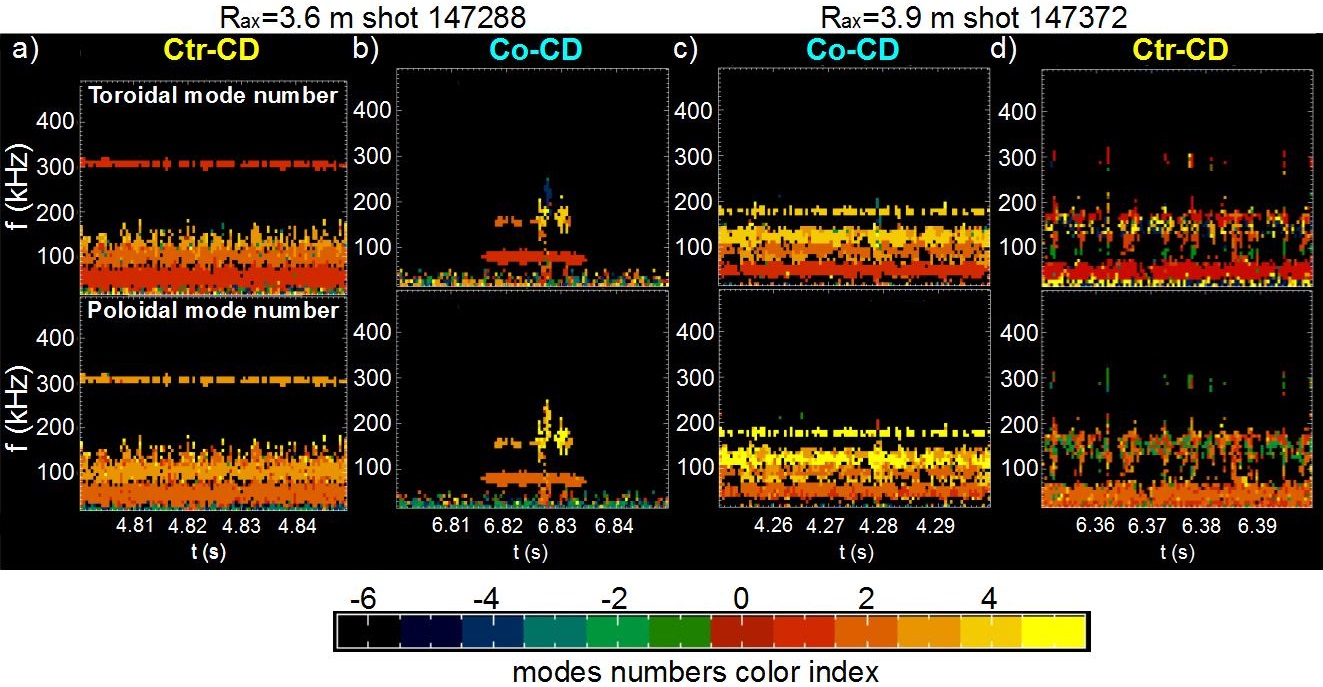}
\caption{Instability mode number measured by the magnetic probe arrays in the discharge $147288$ during the ctr-NBCD (a) and co-NBCD (b) phases. Instability mode number measured by the magnetic probe arrays in the discharge $147372$ during the co-NBCD (c) and ctr-NBCD (d) phases.}\label{FIG:20}
\end{figure*}

Figure~\ref{FIG:20} shows the mode number of the instabilities observed during the co- and ctr-NBCD phases of the discharges $147288$ and $147372$ obtained from the measurement of the magnetic probe arrays.

The ctr-NBCD phase of the discharge $147288$ (Fig.~\ref{FIG:20}a) shows $n/m = 1/2$ instabilities with $f < 80$ kHz, $2/3$ with $f = 100$ kHz, $3/4$ with $f=140$ kHz and $1/3$ with $f = 300$ kHz. The simulations (Fig.~\ref{FIG:9}b) indicate unstable $n=1$ modes with $f=100-120$ kHz and several marginally stable $n=1$ modes with $f < 100$ kHz. Also, $n=2$ and $n=3$ modes are marginally stable in the range of $100-115$ and $160$ kHz, respectively. The marginally stable $n=1$ to $n=3$ modes in the simulations with balanced NBCD should be unstable in a hypothetical ctr-NBCD case as the stability trend indicates, thus the code results are consistent with the experiment. The co-NBCD phase of the discharge $147288$ (Fig.~\ref{FIG:20}b) shows $n/m = 1/2$ instabilities with $f = 30 - 90$ kHz as well as $2/3$ and $3/5$ with $f > 100$ kHz. The simulations (Fig.~\ref{FIG:9}f) indicate unstable $n=1$ modes with $f = 40$ and $125$ kHz, $n=2$ with $f = 60$ and $120$ kHz, $n=3$ with $f =75 – 100$ kHz and $n=4$ with $105$ kHz, consistent with the diagnostic data.

The co-NBCD phase of the discharge $147372$ (Fig.~\ref{FIG:20}c) shows $n/m = 1/2$ instabilities with $f = 50$ kHz, $2/3$ with $f = 100$ kHz, $3/5$ with $f=140$ kHz and $f = 190$ kHz. The simulations (Fig.~\ref{FIG:9}h) indicate unstable $n=1$ modes with $f = 45$ kHz, $n=2$ with $f = 95$ and $135$ kHz, $n=3$ with $f = 110$ and $140$ kHz and $n=4$ with $f = 165$ kHz, similar stability trends compared to the experiment. The ctr-NBCD phase of the discharge $147372$ (Fig.~\ref{FIG:20}d) shows $n/m = 1/2$ instabilities with $f < 60$ and $150$kHz, $2/2$ with $f = 120$ kHz and several $n=3-4$ with $f < 30$ and $f=130-200$ kHz. The simulations (Fig.~\ref{FIG:9}d) indicate unstable $n=1$ modes with $f=55-65$ and $110-115$ kHz, $n=2$ with $f=145$ kHz and several $n=3$ particularly for $f = 10-20$ kHz, $50$ kHz and $110-160$ kHz, also consistent with the magnetic probe measurements.

It should be noted that the poloidal number identification by the present simulations is inaccurate because the model uses an analytic expression for the EP density profile. A correct poloidal number identification requires simulations using EP density profiles calculated by codes as TRANSP \cite{103} or MORH \cite{104,105} for each discharge.

The radial location and mode structure of the instability can be analyzed using the electron cyclotron data (ECE) and the electron density fluctuations (EDF) in LHD, although this data fails to reproduce the AE profiles for these discharge, thus no possible comparison can be done with the simulation results. The analysis performed using the FAR3d code for DIII-D discharges with reverse magnetic shear showed a reasonable agreement between the radial location and structure of the mode eigenfunction calculated in the simulations with the ECE and electron temperature fluctuations data obtained during the experiments \cite{106}.

\ack

The authors wish to thank the LHD experiment and technical staff for their contributions in the operation and maintenance of LHD. This work is supported in part by NIFS under contract NIFS07KLPH004. The authors also wish to acknowledge K. Nagasaki, S. Yamamoto, Y. Suzuki and K. Ida for fruitful discussion.

\hfill \break

\end{document}